\documentclass[10pt,journal,twoside,final,]{IEEEtran}
\usepackage{cite}
\usepackage{graphicx}
\usepackage{times}
\usepackage{latexsym}

\usepackage[mathscr]{eucal}
\usepackage{array}
\usepackage{fancyhdr}
\usepackage{amsmath,amssymb}
\usepackage{bm} 
\usepackage{subfigure}
\usepackage{citesort}
\usepackage{multirow}

\ifCLASSINFOpdf
\else
\fi

\newcommand{\defeq}{\stackrel{\Delta}{=}}

\newcommand{\vect}[1]{{\lowercase{\mathbf{#1}}}}
\newcommand{\mat}[1]{{\uppercase{\mathbf{#1}}}}

\newcommand{\diag}{{\rm{diag}}}
\renewcommand{\d}{\vect{d}} % . accent

\newcommand{\p}{\vect{p}}
\newcommand{\A}{\mat{A}}
\newcommand{\B}{\mat{B}}
\newcommand{\C}{\mat{C}}

\newcommand{\I}{\mat{I}}

\newcommand{\M}{\mat{M}}

\renewcommand{\P}{\mat{P}}

\usepackage{xcolor}

\makeatother

\usepackage{amsthm}
\usepackage{amssymb}
\usepackage{graphicx}
\usepackage{subfigure}
\usepackage{tabularx}
\usepackage{color,cite}
\usepackage{booktabs}
\usepackage{url}
\usepackage{bm}
\usepackage{color}
\definecolor{c}{rgb}{1,0,0} % red
\definecolor{b}{rgb}{0,0,1} % red

\usepackage{algpseudocode}
\usepackage{algorithm}

\usepackage{makecell}
\usepackage{amsfonts}

\usepackage{graphicx}
\usepackage{float}
\usepackage[caption=false,font=footnotesize]{subfig}

\newtheorem{remark}{Remark}
\newtheorem{theorem}{Theorem}
\newtheorem{lemma}[theorem]{Lemma}

\usepackage{algorithm}

\usepackage{algpseudocode}
\begin{document}

\title{Reconfigurable Intelligent Surface-Assisted Secret Key Generation in Spatially Correlated Channels}

\author{
	Lei~Hu,~\IEEEmembership{Student Member,~IEEE},
	Guyue~Li,~\IEEEmembership{Member,~IEEE},
		Xuewen Qian, Aiqun~Hu,~\IEEEmembership{Senior Member,~IEEE}, and  Derrick Wing Kwan Ng,~\IEEEmembership{Fellow,~IEEE}
	\thanks{Part of this paper has been accepted by the IEEE Globecom 2022~\cite{22Globecom}.
		 (Corresponding author: Guyue Li.)}
	\thanks{Lei Hu and Guyue Li are with the School of Cyber Science and Engineering, Southeast University, Nanjing 210096, China. Guyue Li is also with Purple Mountain Laboratories, Nanjing 211111, China, and also with the Jiangsu Provincial Key Laboratory of Computer Network Technology, Nanjing 210096, China (e-mail: lei-hu@seu.edu.cn; guyuelee@seu.edu.cn.).}
	
	\thanks{Xuewen Qian is with with Université Paris-Saclay, CNRS,
		CentraleSupélec, Laboratoire des Signaux et Systèmes, 91192 Gif-sur-Yvette,
		France (e-mail: xuewen.qian@centralesupelec.fr).}
	\thanks{Aiqun Hu is with the National Mobile Communications Research Laboratory, Southeast University, Nanjing, 210096, China, also with Purple Mountain Laboratories, Nanjing 211111, China, and also with the Jiangsu Provincial Key Laboratory of Computer Network Technology, Nanjing 210096, China (e-mail: aqhu@seu.edu.cn).} 
	
	\thanks{Derrick Wing Kwan Ng is with the School of Electrical Engineering and Telecommunications, University of New South
		Wales, Sydney, NSW 2052, Australia (e-mail: w.k.ng@unsw.edu.au).} 
}

\maketitle

\begin{abstract}
Reconfigurable intelligent surface (RIS) is a disruptive technology to enhance the performance of physical-layer key generation (PKG) thanks to its ability to smartly customize the radio environments. Existing RIS-assisted PKG methods are mainly based on the idealistic assumption of an independent and identically distributed (i.i.d.) channel model at both the base station (BS) and the RIS. However, the i.i.d. model is inaccurate for a typical RIS in an isotropic scattering environment and neglecting the existence of channel spatial correlation would possibly degrade the PKG performance. In this paper, we establish a general spatially correlated channel model and propose a new channel probing framework based on the transmit and the reflective beamforming. We derive a closed-form key generation rate (KGR) expression and formulate an optimization problem, which is solved by using the low-complexity Block Successive Upper-bound Minimization (BSUM) with Mirror-Prox method. Simulation results show that compared to the existing methods based on the i.i.d. fading model, our proposed method achieves about $5$ dB transmit power gain when the spacing between two neighboring RIS elements is a quarter of the wavelength. Also, the KGR increases significantly with the number of RIS elements while that increases marginally with the number of BS antennas.
\end{abstract}


\section{Introduction}
In the last decades, 
the throughput of wireless communication systems has 
achieved a 1000-fold capacity increase~\cite{ylianttila20206g}.
 At the same time, an enormous amount of confidential information, including financial information and trade secrets, has been exchanged via wireless channels.
However, the broadcast property of wireless medium makes wireless transmissions vulnerable to security breaches, such as passive and active attacks by potential eavesdroppers. 
Traditionally, security communication
is guaranteed by the public key cryptography techniques adopted in the application layer. 
However, the public key in those traditional methods needs to be distributed to the involved legitimate ends in advance, which is difficult to realize in mobile and ad-hoc wireless networks. 
%Alternatively, physical-layer security (PLS) s are composed of keyless security and secret key-based secrecy
Alternatively, physical-layer key generation (PKG) is recognized as a promising paradigm to tackle this problem~\cite{JiaoWCM}. By leveraging the inherent random and reciprocal nature of wireless channels, PKG naturally establishes symmetric keys between the
legitimate parties \cite{maurer1993secret}. Also, due to
its potential to achieve information-theoretic security cost-effectively, PKG has been applied to practical systems, such as WiFi, LoRa, and Zigbee, {\it et al}~\cite{ZhangReview},\cite{mathur2008radio}.

The general PKG procedures can be divided into four consecutive stages: channel sounding, quantization, information reconciliation, and privacy amplification \cite{li2018high}. Specifically, for the channel probing stage, Alice and Bob alternatingly exchange pilots and perform channel estimations in a coherence time slot to extract correlated channel features. These features are converted into binary bits in the quantization stage. Then, in the information reconciliation stage, the mismatched bits between Alice and Bob are corrected via error-correcting codes. At last, during the privacy amplification stage, Alice and Bob discard the bits that could be potentially leaked in the previous stages. 
It can be observed that the feasibility of the PKG is strongly associated with the properties of wireless channels. Moreover, the existence
	 of rich-scattering and dynamically varying channels is the essential premise to
ensure the security level offered by the generated secret key~\cite{ZhangReview},\cite{2021Encrypting}.

Unfortunately, the above condition can hardly be satisfied in some harsh propagation scenarios, such as shadowed environments~\cite{Li2022WCM}. 
As such, various previous works aim to enhance the PKG performance in these propagation scenarios to a certain extent by utilizing cooperative relaying~\cite{6716049}. However, the relay-based PKG approaches admit two main demerits. First, the key
rate has limited growth unless the relay node
keeps moving all the time to introduce randomness to the secret key~\cite{Gui2015Untrusted}.
Second, deploying
active relays incurs
high hardware costs and energy consumption \cite{YuGreen}.
As a result, a new cost-effective and energy-effective paradigm that is capable of controlling the propagation environment
is needed to facilitate PKG.

%For example, in wireless sensor networks (WSN), the sensor nodes are static or with little movement. Hence the channel is slow-varying and the secret keys cannot be updated frequently.
%In addition, in a factory automation scenario where
%equipments are densely deployed, the radio wave may be
%blocked by obstacles, leading to an interruption of communication
%as well as the secret key generation. 
%As a result, there is a need for new technologies to overcome these two
%demerits. 

As a remedy, reconfigurable intelligent surface (RIS) was proposed to address this emerging need. In particular, RIS is a programmable and reconfigurable metasurface consisting of a large number of low-cost passive elements, e.g., printed dipoles and
phase shifters~\cite{di2019smart},\cite{Smart}. These elements are controlled collaboratively to alter the signal propagation environment. Hence, RIS provides
a cost-effective approach to enable the customization
of favorable wireless propagation environments for PKG while avoiding the deployment of power-hungry and expensive cooperative relays. 
%The works \cite{21JiWCL,2020Intelligent,kai_RIS,2022lu} investigated the RIS-induced randomness method to generate high-entropy key in quasi-static environments. To be specific, in \cite{21JiWCL}, the authors derived the  theoretical expression of the key generation rate (KGR), where the phase shifting of RIS elements follows discrete random distribution. They also designed an optimal time slot allocation algorithm to maximize the encrypted data transmission rate. In addition, 
%the authors in \cite{kai_RIS} proposed a machine
%learning based adaptive quantization level prediction scheme to achieve high KGR while keeping low BDR.
%%In above works, different sub-reflecting channels, that introduced by RIS elements, are regarded to be superimposed. Hence the entropy of the RIS-introduced is reduced. 
%In \cite{2022lu}, the authors proposed to exploit the randomness of the sub-reflecting channels to fully exploit the randomness of RIS. The lower and upper bounds of the KGR are proved to be further improved. 
%Furthermore, a RIS-based PKG prototype was implemented by using the commodity Wi-Fi transceivers in \cite{2020Intelligent}.  The experimental results verified the effectiveness of the RIS in improving the randomness of the secret key.
However, 
%to reap the aforementioned benefits promised of the RIS, 
to fully unleash the potential of RIS for effective KGR provisioning,
the reflection coefficients of RIS elements have to be appropriately optimized.
% needs to be designed to fully unleash the potential of RIS for KGR provisioning.
Inspired by this, several works have been proposed on the optimization design of RIS-assisted PKG systems, such as [15]--[17],
% \cite{2021SPL,21JiTVT,22Sum-RIS}, 
starting from single-user systems to multi-user systems. In particular, the authors of \cite{2021SPL} considered a simple scenario with a
single-user and independent eavesdropping channels. They derived the expression of the KGR capacity and optimized the switch states of RIS units globally when the number of available RIS units is limited. 
Furthermore, RIS-assisted PKG systems comprising multiple eavesdroppers with correlated eavesdropping channels were investigated in \cite{21JiTVT}, where the authors designed a semidefinite relaxation (SDR) and successive
convex approximation (SCA)-based algorithm to maximize the lower bound of the secret key capacity.
Despite the fruitful research in the literature, most of the above works considered only the single-user case, e.g., \cite{2021SPL}, \cite{21JiTVT}, and their results are not applicable to practical cases having multiple users. 
On the other hand, in \cite{22Sum-RIS}, the authors introduced a RIS-assisted multiuser key generation scheme and
optimized the RIS phase shifts in the presence of independent and correlated eavesdropping channels.
%. Specifically, in the presence of independent eavesdropping channels, they found the optimal RIS phase shifts based on the Karush-Kuhn-Tucker condition. In the presence of correlated eavesdropping channels, a SDR-SCA based algorithm was employed to optimize the RIS phase shifts.
Nevertheless, all of these works
are based on the independent and identically distributed (i.i.d.)
Rayleigh fading model for the RIS-related channels. In practice,
the non-negligible spatial correlations naturally exist among RIS
elements due to their sub-wavelength sizes and distributions. More importantly,
these correlations may jeopardize the PKG performance if
they are not taken into account in the system design \cite{Rayleigh-fading}.
In addition, only a single-antenna BS was considered in
these works, e.g.,~[15]--[17], and it is not straightforward to extend
these existing results to the case of multi-antenna, which has a more
complicated model, and the reflective beamforming at RIS and transmit beamforming at the BS need to be jointly optimized. 
Indeed, the design of RIS-assisted PKG methods in multi-antenna spatially correlated channels is of utmost importance but it has not been studied in the existing works, yet.

To address the above issues, this paper investigates the RIS-assisted PKG in a 
multiple-input single-output multiple-eavesdropper 
(MISOME) system, with the consideration of the spatial correlation at both the BS and the RIS. 
%We derive the KGR expression and present an effective algorithm to jointly optimize the transmit beamforming and reflective beamforming.
The main contributions of this paper are as follows.
\begin{itemize}
	\item We propose a transmit and reflective beamforming-based RIS-assisted PKG framework in multi-antenna spatially correlated channels. 
	We derive the closed-form KGR expression as a function of the reflective beamforming at the RIS and the transmit beamforming at the BS. We formulate the design of beamforming as an optimization problem to maximize the minimum KGR for the
	worst-case eavesdropper channel.	
	Furthermore, our analysis shows that
%	we analyze the KGR performance difference
%		between the one adopting the assumption of the i.i.d. model
%		and that of the spatially correlated model. 
%	It is found that 
	the
	beamforming designed for the correlated model outperforms that
	for the i.i.d. model while the KGR gain increases with the channel
	correlation with the proposed design.

	\item To tackle the resulting non-convex optimization problem, we present an effective Block Successive Upper-bound Minimization
	(BSUM)-based algorithm. 
	We prove that the BSUM algorithm yields a non-decreasing convergence over iterations.
%	 The required number of iterations to convergence is small in simulations, which reduces the computation cost. 
	Then, to solve the non-smooth convex problem in each iteration of the BSUM algorithm in a complexity-effective manner, we reformulate it as an equivalent convex-concave saddle point problem and employ the Mirror-Prox method to solve it with closed-form updates. 
%	In each iteration of the proposed method, the solution is in closed form, and thus the computation complexity is low.
	
	\item Simulation results show that 
	compared to existing
	methods based on the i.i.d. fading model, the proposed
	design achieves about $5$ dB transmit power  gain when the spacing between two neighbouring RIS elements is a quarter of the wavelength and the BS antenna correlation $\rho$ is 0.2. Also, the KGR gain increases with
	the spatial correlation at both the BS and the RIS.
	Moreover, the computational time of the proposed
	algorithm is reduced approximately by two orders of magnitude compared to that of the
	commonly adopted algorithms, e.g., alternating optimization, semidefinite relaxation, successive convex approximation with Gaussian randomization (ASSG), while achieving a similar KGR performance. 
\end{itemize}

%The material in this paper has been partially submitted to IEEE Globecom 2022~. 
Note that in the conference version of this paper~\cite{22Globecom}, we considered the case where 
%presented the transmit and reflective beamforming based PKG framework and derived the KGR when 
the eavesdropping channels are independent of the legitimate
channels and derived the optimal beamforming to maximize the KGR.
In this paper, we generalize it by taking into account the channel correlation between Eve and the legitimate parties. We provide a general KGR expression and formulate an optimization problem, which can be solved by the proposed low-complexity BSUM algorithm.

%\subsection{Paper Organization and Notations}
%The rest of the paper is organized as follows. Section~II describes the spatially correlated channel model and the transmit and reflective beamforming based PKG framework. In Section~III, we give the KGR expression and formulate an optimization problem. Section~IV presents an DBMM framework to tackle the optimization problem. Section~V illustrates the Mirror-Prox method to solve the subproblems in each DBMM iteration. 
%Section~V gives the impact of different beamforming methods on PKG. 
%The simulation results are provided in Section VII. Finally, we conclude this paper in Section~VIII.

\emph{Notations:} 
In this paper, matrices and vectors are denoted by boldface upper-case and boldface lower-case. 
$\mathbb{C}^{A\times B}$ denotes the space of complex matrices of size $A\times B$. 
$\Re(\cdot)$ and $\Im(\cdot)$ stand for the real and imaginary parts of a complex number. The imaginary unit of a complex number is denoted by $j=\sqrt{-1}$.
$(\cdot)^*$, $(\cdot)^{\sf T}$, and $(\cdot)^{\sf H}$ denote the conjugate, transpose, and conjugate transpose, respectively. $\diag(\boldsymbol{x})$
is a matrix whose main diagonal elements are the entries of $\boldsymbol{x}$.
 $\text{vec}(\boldsymbol{X})$ denotes the vectorization of the matrix $\boldsymbol{X}$. 
% $\tr(\cdot)$ and $\text{rank}(\cdot)$ represent the trace and rank of a matrix. 
% $\triangleq$ means “defined as”. 
 The Kronecker product, Hadamard product, and Khatri-Rao product are represented by $\otimes$, $\circ$, and $\odot$, respectively. $\mathcal{I}(X;Y)$ and $\mathcal{H}(X,Y)$ are the mutual information and joint entropy of random variables $X$ and $Y$, respectively. 
%$\mathcal{H}(X,Y|Z)$ is the conditional entropy of $X$ and $Y$ given $Z$.
$\operatorname{det}(\cdot)$ is the matrix determinant. $||\boldsymbol{x}||_{1}$, $||\boldsymbol{x}||_{2}$, and $||\boldsymbol{x}||_{\infty}$ denote the $\ell_{1}$, $\ell_{2}$, and $\ell_{\infty}$ norms of vector $\boldsymbol{x}$. $\|\boldsymbol{X}\|_{F}$ is the Frobenius norm of matrix $\boldsymbol{X}$.
$\mathbb{E}\{\cdot\}$ represents statistical expectation. 
$\lambda_{\mathrm{max}}(\boldsymbol{X})$ is the maximum eigenvalue of matrix $\boldsymbol{X}$.
$\mathcal{O}(\cdot)$ is the big-O notation.
$\boldsymbol{X} \succeq \boldsymbol{0}$ means $\boldsymbol{X}$ is a positive semidefinite matrix.
$\nabla f(\cdot)$ and 
$\partial f/\partial x$ are the gradient operators of function $f$. $\boldsymbol{I}$ denotes the identity matrix. 
$\boldsymbol{b}\sim \mathcal{CN}\left(\boldsymbol{0}, \boldsymbol{\Sigma}\right) $ denotes that $\boldsymbol{b}$ is a circularly
symmetric complex Gaussian (CSCG) vector with zero mean and
covariance matrix $\boldsymbol{\Sigma}$.

\section{System Model}\label{sec:system_model}
\begin{figure}
	\centering
	\includegraphics[width=0.45\textwidth]{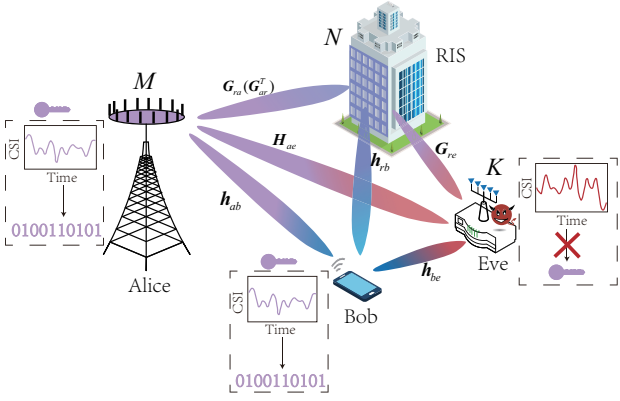}
	\caption{The model of RIS-assisted secret key generation in MISOME systems.}
	\label{system model}
\end{figure}
As shown in Fig.~\ref{system model}, we study a RIS-assisted PKG method in a MISOME system. Assuming the time-division duplexing (TDD) protocol is adopted, a multi-antenna base station (BS), Alice, and a single-antenna user, Bob, aim to generate symmetric keys by exploiting the reciprocity of the wireless channel with the help of a RIS\footnote{The case studied in this paper can be directly extended to multi-users scenarios, since the pilots used by different users are orthogonal to each other in a single-cell system~\cite{Yin-pilot} and thus their PKG processes are independent.}. Meanwhile, a multi-antenna eavesdropper, Eve, intends to obtain the secret key information from his/her received signals. 
%Specifically, assuming a time-division duplexing (TDD) mode, Alice and Bob sound channel alternatively in coherence time to obtain the reciprocal channel estimation.

We assume that Alice and Eve are equipped with $ M $ and $ K $ antennas, respectively. The RIS consists of $N$ passive reflection elements. Equipped with a smart controller that communicates with the BS, the RIS adapts the phase shift of
each reflecting element to enable the secret key generation \cite{wu2019towards}. 
%Contrary to existing works, relying on independent Rayleigh channel models, we consider correlated Rayleigh fading
Since the spatial correlation affects the secret key rate, we consider the general spatial correlation channel model at both the RIS and the BS.

\subsection{Channel Model}

The direct channels of Alice-to-Bob, Eve-to-Bob, and Alice-to-Eve are denoted by $\boldsymbol{h}_{ab}  \in \mathbb{C}^{M \times 1}$, 
$\boldsymbol{h}_{eb}\in \mathbb{C}^{K \times 1}$, and $\boldsymbol{H}_{ae}  \in \mathbb{C}^{M \times K}$, respectively. $\boldsymbol{h}_{ak} \in \mathbb{C}^{M \times 1}$ denotes the channel from Alice to Eve's $k$-th antenna, $k\in \{1,\cdots,K\}$. 
When a RIS is involved in the PKG system, it introduces additional communication channels.
Specifically, the channels of RIS-to-Alice, RIS-to-Bob, and RIS-to-Eve are represented as $\boldsymbol{G}_{ra}\in \mathbb{C}^{N \times M}$, $\boldsymbol{h}_{r b}\in \mathbb{C}^{N \times 1}$, and $\boldsymbol{G}_{r e} \in \mathbb{C}^{N \times K}$, respectively.  $\boldsymbol{h}_{r k} \in \mathbb{C}^{N \times 1}$ denotes the channel from the RIS to Eve's $k$-th antenna.
%where $ \boldsymbol{h}_{kr} $ for $k = 1, \cdots, K$ in $ \boldsymbol{H}_{e r} $ is the channel from Eve's $k$-th antenna to RIS. 
%The direct channels $ \boldsymbol{h}_{ba} \in \mathbb{C}^{M \times 1} $, $ \boldsymbol{h}_{be}=[h_{b1}, \cdots,h_{bK}]^{T} \in \mathbb{C}^{K \times 1} $, and $ \boldsymbol{H}_{ea}=[\boldsymbol{h}_{1,a},\cdots,\boldsymbol{h}_{K,a} ] \in \mathbb{C}^{M \times K} $ denote the channel from Bob to Alice, from Bob to Eve, and from Eve to Alice, respectively. 
To account for the spatial correlation, the channel matrices are described by employing the Kronecker correlation channel model as \cite{yang2020asymptotic} 
\begin{align}
	\boldsymbol{G}_{ar} &=\boldsymbol{G}_{ra}^{\sf T}= \beta_{ra}^{\frac{1}{2}}
	\boldsymbol{R}_{S}^{\frac{1}{2}} \tilde{ \boldsymbol{H}} \boldsymbol{R}_{I}^{\frac{1}{2}}
	, \label{eq1}\\
	\boldsymbol{h}_{ri}&=  \beta_{ri}^{\frac{1}{2}}\boldsymbol{R}_{I}^{\frac{1}{2}} \tilde{ \boldsymbol{h}}_{ri}, \ i \in \{b, k\}, \label{eq2} \\
	\boldsymbol{h}_{aj}&= \beta_{aj}^{\frac{1}{2}}\boldsymbol{R}_{S}^{\frac{1}{2}} \tilde{ \boldsymbol{h}}_{aj}, \ j \in \{b,k\}, \label{eq3}
\end{align}
respectively, where $\boldsymbol{R}_{S} \in \mathbb{C}^{M \times M}$ and $\boldsymbol{R}_{I} \in \mathbb{R}^{ N \times N} $ are the spatial correlation matrices at Alice and the RIS, respectively\cite{Rayleigh-fading}.
% $\tr(\cdot)$ denotes the trace of a matrix. 
%The elements $\left[ \boldsymbol{R}_{S}\right]_{m,n}$ and $\left[ \boldsymbol{R}_{I}\right]_{m,n}$ represent the correlation between the $m$-th antenna/element and the $n$-th antenna/element. 
In addition, 
$\tilde{ \boldsymbol{H}} \in \mathbb{C}^{M \times N}$, $\tilde{ \boldsymbol{h}}_{ri} \in \mathbb{C}^{N \times 1}$, and $\tilde{ \boldsymbol{h}}_{aj} \in \mathbb{C}^{M \times 1}$ are random matrices with i.i.d. Gaussian random entries of zero mean and unit variance. 
$ \beta_{ra}$, $\beta_{ri}$, and $\beta_{aj}$ are the path loss of the corresponding channels, respectively.

\subsection{PKG Framework Based on Transmit and Reflective Beamforming} 
%In PKG system, Alice and Bob treat the reciprocal wireless channel as common randonmess to extract a symmetric key. 
%The general key generation process consists four steps, i.e. , while in this paper we focus on the first step to maximize the theoretical key rate. 
Now, we propose a new framework to take full advantage of the RIS-assisted PKG in multi-antenna systems. 
%In this paper, 
%Take full advantage of the RIS-assisted PKG in multi-antenna systems,
%%Different from the SISO system that only the RIS coefficients are configured to improve the key generation rate, 
%we propose to joint design the transmit and reflective beamforming at BS and RIS, respectively. 
%The general PKG contains four steps, namely channel probing, quantization, information reconciliation, and privacy amplification. Since the last three steps are similar to existing works [access], in this paper we focus on the channel probing step where transmit and reflective beamforming are joint designed to facilitate PKG. 
In the PKG system, Alice and Bob first perform channel probing to acquire the reciprocal channel estimation. 
The process of channel probing is described as follows.

\emph{Step 1: Uplink channel sounding.} Bob transmits the publicly known pilot $s_{u} \in\mathbb{C} $ with $|s_{u}|^2=1$. Then, the equivalent baseband signal received at Alice and Eve is expressed as 
\begin{align}
	\boldsymbol{y}_{l}^{u} =  \sqrt{P_\mathrm{B}} \left(\boldsymbol{G}_{lr}  \boldsymbol{\Phi}\boldsymbol{h}_{rb} + \boldsymbol{h}_{lb}    \right) s_{u} + \boldsymbol{z}_{l}, l\in \{a,e\},
\end{align}
respectively, where $\boldsymbol{\Phi} = \diag\{\boldsymbol{v}\}$ with each element $|v_{n}|\leq 1, \forall n\in  \{1,\cdots,N\}$, is the reflection coefficients matrix of the RIS. In addition, $P_\mathrm{B}$ is the transmit power of Bob. The noise follows the circularly symmetric complex Gaussian distribution, i.e., 
$\boldsymbol{z}_{l} \sim \mathcal{CN}(0, \sigma_{l}^2 \boldsymbol{I})$, with $\sigma_{ l }^2$ being
the corresponding noise variances. 
Then, Alice and Eve perform a standard least-squares (LS) channel estimation~\cite{21JiTVT},\cite{22Sum-RIS} as\footnote{The LS is adopted since it has been widely used in practical systems\cite{2020Intelligent}.}
\begin{align}
		\hat{\boldsymbol{h}}_{l}^{u} \triangleq   s_{u}^*\boldsymbol{y}_{l}^{u} 
		= \sqrt{P_\mathrm{B}} (\boldsymbol{G}_{lr} \boldsymbol{\Phi}\boldsymbol{h}_{rb} +   \boldsymbol{h}_{lb}) + \tilde{\boldsymbol{z}}_{l}^{u}, l\in \{a,e\}, \label{ha1} 
\end{align} 
respectively, where the estimation noise is $ \tilde{\boldsymbol{z}}_{l}^{u}=s_{u}^*\boldsymbol{z}_{l}^{ u }  $ and $\hat{\boldsymbol{h}}_{e}^{u}=[\hat{{h}}_{e_{1}}^{u},\cdots, \hat{{h}}_{e_{K}}^{u}]^{\sf T}$. 

\emph{Step 2: Downlink channel sounding.} Alice sends the downlink publicly known pilot $s_{d} \in \mathbb{C}$ with $|s_{d}|^2 = 1 $ and the signals received at Bob and Eve are 
\begin{align}
	y_{b}^{d} &=  (\boldsymbol{h}_{rb}^{\sf T} \boldsymbol{\Phi} \boldsymbol{G}_{ra}  +  \boldsymbol{h}_{ab}^{ \sf T } )\boldsymbol{w} s_{d} +  z_{b}^{d}, \\
	\boldsymbol{y}_{e}^{d} &=  (\boldsymbol{G}_{re}^{\sf T} \boldsymbol{\Phi} \boldsymbol{G}_{ra}  +  \boldsymbol{H}_{ae}^{\sf T} )\boldsymbol{w} s_{d} +  \boldsymbol{z}_{e}^{d}, 
\end{align}
respectively, where $\boldsymbol{w}\in\mathbb{C}^{M\times 1}$ is the transmit beamforming vector at Alice that satisfies $||\boldsymbol{w}||_{2}^2 \leq P_\mathrm{A}$ with $P_\mathrm{A}$ being the maximum transmit power of Alice. 
$z_{b}^{d}$ and $\boldsymbol{z}_{e}^{d} $ are the additive Gaussian noise with $z_{b}^{d} \sim \mathcal{CN}\left(0, \sigma_{b}^2\right)$ and $\boldsymbol{z}_{e}^{d} \sim \mathcal{CN}\left(0, \sigma_{e}^2\boldsymbol{I}\right)$. 
After the LS estimation, Bob and Eve obtain the channel estimates as 
\begin{align}
		\hat{ h}_{ b }\triangleq  s_{d}^*y_{b}^{d} 
		&= (\boldsymbol{h}_{rb}^{\sf T}\boldsymbol{\Phi} \boldsymbol{G}_{ra}  +  \boldsymbol{h}_{ab}^{ \sf T } )\boldsymbol{w} + \tilde{z}_{b},  \label{hb} \\
		\hat{\boldsymbol{h}}_{ e }^{d} \triangleq  s_{d}^* \boldsymbol{y}_{e}^{d} 
		&= (\boldsymbol{G}_{re}^{\sf T} \boldsymbol{\Phi} \boldsymbol{G}_{ra}  +  \boldsymbol{H}_{ae}^{\sf T} )\boldsymbol{w} + \tilde{\boldsymbol{z}}_{e}^{u},
	\end{align}
	respectively, where the noises are $\tilde{z}_{b}^{ d } = s_{d}^* z_{b}^{d} $ and $\tilde{\boldsymbol{z}}_{e}^{u}=s_{d}^*\boldsymbol{z}_{e}^{u}$, respectively, and $\hat{\boldsymbol{h}}_{ e }^{d} = [\hat{h}_{ e_{1} }^{d},\cdots,\hat{{h}}_{ e_{K} }^{d}]^{\sf T}$.

\emph{Step 3: Reciprocal components acquisition.} 
Since the estimations obtained by Alice and Bob, as shown in (\ref{ha1}) and (\ref{hb}), are quite different in terms of both the dimensions and values, we multiply Alice's channel estimation $ \hat{\boldsymbol{h}}_{a}^{u}$ by $\boldsymbol{w}$ to obtain the combined reciprocal channel gain as
	\begin{align}
		\hat{ h}_{a}  \triangleq  \boldsymbol{w}^{\sf T}  \hat{\boldsymbol{h}}_{a}^{u}= \sqrt{P_\mathrm{B}} \boldsymbol{w}^{\sf T} ( \boldsymbol{G}_{ar} \boldsymbol{\Phi}\boldsymbol{h}_{rb}+   \boldsymbol{h}_{ab} )+ z_{a}, \label{ha}
	\end{align}
	where the noise is $z_{a}=\boldsymbol{w}^{\sf T} \tilde{\boldsymbol{z}}_{a}^{u}$. 

Consequently, Alice's combined channel gain, $\hat{ h}_{a}$, and Bob's channel gains, $\hat{ h}_{b}$, are highly correlated in general. 
After following the procedures of the PKG, i.e., quantization, information reconciliation, and privacy amplification, the channel gains are finally converted into secret keys \cite{2021Sum}. Since
these steps are similar to those adopted in existing PKG methods, such as \cite{li2018high}, \cite{ZhangReview}, in this paper, we focus on 
the channel probing step, where the transmit and reflective beamforming are jointly optimized to maximize the KGR.

%Compare the combined channel $\tilde{ h}_{a}$ and $\tilde{ h}_{b}$ in expressions \ref{ha} and \ref{hb}, Alice and Bob obtain recipical component. 
%Then, Alice and Bob will quantizate the channel estimation $\tilde{ h}_{a}$ and $\tilde{ h}_{b}$ respectively. To generate same bits, in information  reconciliation step, error-correcting codes could be employed to correct bits disagreements. At last, Privacy amplification ... [TIFS/Access]. 

\section{Problem Formulation and Transformation}
In this section, based on the channel estimation acquired in Sec. \ref{sec:system_model}, 
we derive the closed-form KGR expression and formulate an optimization problem regarding the variables $\boldsymbol{w}$ and $\boldsymbol{v}$ to improve the system performance.
%$\boldsymbol{w}$ and $\boldsymbol{v}$
%of the transmit beamforming $\boldsymbol{w}$ and the reflective beamforming $\boldsymbol{v}$. 

%we derive a closed-form expression for the secret key rate based on the proposed framework in Sec. \ref{Sec-Model}-B. 

% which can be expressed as 
%\begin{align}
%	%	R_{SK}  \triangleq \mathcal{I}\left(\tilde{ h}_{a} ; \tilde{ h}_{b}|  \overline{h}_{e1},\cdots, \overline{h}_{eK},  \tilde{ h}_{ e1 },\cdots, \tilde{ h}_{ eK } \right),
%	R_{\mathrm{SK}}  \triangleq \mathcal{I}\left(\hat{ h}_{a} ; \hat{ h}_{b}|  \hat{\boldsymbol{h}}_{e}^{u}, \hat{\boldsymbol{h}}_{e}^{d} \right),
%\end{align}

%As obtaining the exact secret key
%rate given multi-antenna Eve's channel estimates is still an open problem,
We first formulate the KGR given Eve's $k$-th antenna's channel estimation
and maximize the minimal KGR\footnote{The KGR expression and the beamforming design in this paper can also be applied to the scenario with multiple single-antenna eavesdroppers. }~\cite{21JiTVT}. 
Specifically, the KGR for Eve's $k$-th antenna is defined as the conditional mutual information of the legitimate parties' channel estimations given the observation of Eve's $k$-th antenna \cite{li2018high}, i.e.,
%\begin{align}
	$R_{ k} = \mathcal{I}\left(\hat{ h}_{a} ; \hat{ h}_{b}\mid \hat{ h}_{e_{k}}^{d},\hat{ h}_{e_{k}}^{u}\right)$.
%\end{align}
%Also, Eve is assumed to be a passive eavesdropper, who pretends
%to be an authorized user and intends to infer the keys based on his/her experienced channels~\cite{21JiTVT}.
Since Eve is located close to Bob but locates far away from Alice, the channel estimated by Eve in the uplink channel sounding is statistical independent with the legitimate parties' channel estimations \cite{wallace2009key}. Thus, we have $\mathcal{I}\left(\hat{ h}_{a} ; \hat{ h}_{b}\mid \hat{ h}_{e_{k}}^{d},\hat{ h}_{e_{k}}^{u}\right) = \mathcal{I}\left(\hat{ h}_{a} ; \hat{ h}_{b}\mid \hat{ h}_{e_{k}}^{d}\right)$.
Therefore, the KGR given Eve's $k$-th antenna is given by~\cite{jorswieck2013secret},\cite{wong2009secret}
\begin{align}
	R_{k}
%	=& \mathcal{H}\left(\hat{ h}_{a}, \hat{ h}_{e_{k}}^{d}\right)+\mathcal{H}\left(\hat{ h}_{b}, \hat{ h}_{e_{k}}^{d}\right) \notag \\
%		&-\mathcal{H}\left(\hat{ h}_{a}, \hat{ h}_{b}, \hat{ h}_{e_{k}}^{d}\right)-\mathcal{H}\left(\hat{ h}_{e_{k}}^{d}\right) \\
		= \log_{2} \frac{\operatorname{det}\left(\boldsymbol{R}_{ae_{k}} \boldsymbol{R}_{be_{k}}\right)}{\operatorname{det}\left(\boldsymbol{R}_{abe_{k}} \right)\mathcal{R}_{e_{k}e_{k}}} , \label{eq:MI}
	\end{align}
where the covariance matrices are denoted as
\begin{align}
	\boldsymbol{R}_{ue_{k}} =\left[\begin{array}{ll}
		\mathcal{R}_{uu} & \mathcal{R}_{ue_{k}}  \\
		\mathcal{R}_{e_{k}u } & \mathcal{R}_{e_{k} e_{k}} 
	\end{array}\right], u\in \{a,b\},
\end{align}
%and 
\begin{align}
	\boldsymbol{R}_{abe_{k}} =\left[\begin{array}{lll}
		\mathcal{R}_{a a} & \mathcal{R}_{a b} & \mathcal{R}_{a e_{k}} \\
		\mathcal{R}_{b a} & \mathcal{R}_{b b} & \mathcal{R}_{b e_{k}} \\
		\mathcal{R}_{e_{k} a} & \mathcal{R}_{e_{k} b} & \mathcal{R}_{e_{k} e_{k}}
	\end{array}\right],
\end{align}
respectively. $\mathcal{R}_{xy}=\mathbb{E}\left\{\hat{ h}_{x}\hat{ h}_{y}^{\sf H} \right\}, x,y \in \{ a,b,e_{k}\}$ 
denotes the channel covariances of the corresponding channel estimations. 

Substituting the channel estimations into (\ref{eq:MI}) and assuming $\sigma_{ a }^2=\sigma_{ b }^2=\sigma_{ e_{k} }^2=\sigma^2,k\in\{1,\cdots,K\}$ for simplicity~\cite{21JiTVT},\cite{22Sum-RIS}, we provide the following lemma to characterize the KGR.
\begin{lemma}
	The KGR between Alice and Bob,
	given the channel estimation at Eve's $k$-th antenna, is expressed as (\ref{eq:whole}) at the top of this page,  where
%%写成A+B-C的形式
%\begin{align}
%	g_{u} &= ( \boldsymbol{w}^{\sf T}  \boldsymbol{R}_{S}  \boldsymbol{w}^{ * })(\beta_{ r } \boldsymbol{v}^{\sf H}    \tilde{\boldsymbol{R}}_{I}            \boldsymbol{v} + \beta_{ ba } ), \\
%	g_{ue}^{k} &= (\boldsymbol{w}^{\sf T}  \boldsymbol{R}_{S}  \boldsymbol{w}^{ * })\boldsymbol{v}^{\sf H} \tilde{\boldsymbol{R}}_{bk}^{r}\boldsymbol{v}\sqrt{\beta_{ r } \beta_{ r }^{k}} + \boldsymbol{w}^{\sf T}  \tilde{\boldsymbol{R}}_{bk}^{d}  \boldsymbol{w}^{ * }\sqrt{\beta_{ ba } \beta_{ ka}}, \\
%		g_{e}^{k}&= ( \boldsymbol{w}^{{\sf T}}  \boldsymbol{R}_{S}  \boldsymbol{w}^{ * })(\beta_{ r }^{k} \boldsymbol{v}^{\sf H}   \tilde{\boldsymbol{R}}_{I}          \boldsymbol{v} + \beta_{ ka } ),
%\end{align}
%\begin{align}
	$g_{u} = ( \boldsymbol{w}^{\sf T}  \boldsymbol{R}_{S}  \boldsymbol{w}^{ * })(\beta_{ r } \boldsymbol{v}^{\sf H}    \tilde{\boldsymbol{R}}_{I}            \boldsymbol{v} + \beta_{ ab } )$,
	$g_{ue}^{k} = (\boldsymbol{w}^{\sf T}  \boldsymbol{R}_{S}  \boldsymbol{w}^{ * })\boldsymbol{v}^{\sf H} \tilde{\boldsymbol{R}}_{bk}^{r}\boldsymbol{v}\sqrt{\beta_{ r } \beta_{ r }^{k}} + \boldsymbol{w}^{\sf T}  \tilde{\boldsymbol{R}}_{bk}^{d}  \boldsymbol{w}^{ * }\sqrt{\beta_{ ab } \beta_{ ak}}$, 
	$g_{e}^{k}= ( \boldsymbol{w}^{{\sf T}}  \boldsymbol{R}_{S}  \boldsymbol{w}^{ * })(\beta_{ r }^{k} \boldsymbol{v}^{\sf H}   \tilde{\boldsymbol{R}}_{I}          \boldsymbol{v} + \beta_{ ak } )$,
%\end{align}
$\tilde{\boldsymbol{R}}_{bk}^{r} = ((\boldsymbol{R}_{I}^{\frac{1}{2}}) ^{\sf T }\boldsymbol{R}_{bk}^{r}(\boldsymbol{R}_{I}^{\frac{1}{2}}) ^{\sf T }) \circ   \boldsymbol{R}_{I}$, $\tilde{\boldsymbol{R}}_{bk}^{d}  =  \boldsymbol{R}_{S}^{\frac{1}{2}} \boldsymbol{R}_{bk}^{d}\boldsymbol{R}_{S}^{\frac{1}{2}}$, and $\tilde{\boldsymbol{R}}_{I}=\boldsymbol{R}_{I}^{\sf T}  \circ   \boldsymbol{R}_{I}$.
%\begin{align}
%	\tilde{\boldsymbol{R}}_{bk}^{r} &= ((\boldsymbol{R}_{I}^{\frac{1}{2}}) ^{\sf T }\boldsymbol{R}_{bk}^{r}(\boldsymbol{R}_{I}^{\frac{1}{2}}) ^{\sf T }) \circ   \boldsymbol{R}_{I}, \\
%	\tilde{\boldsymbol{R}}_{bk}^{d} & =  \boldsymbol{R}_{S}^{\frac{1}{2}} \boldsymbol{R}_{bk}^{d}\boldsymbol{R}_{S}^{\frac{1}{2}},\\
%	\tilde{\boldsymbol{R}}_{I}&=\boldsymbol{R}_{I}^{\sf T}  \circ   \boldsymbol{R}_{I}.
%\end{align}
Also, $\beta_{ r }=\beta_{ra}\beta_{rb}$, $\beta_{ r }^{k}=\beta_{ra}\beta_{rk}$, and
the covariance matrices are defined as
$\boldsymbol{R}_{bk}^{r} \defeq \mathbb{E}\{\tilde{ \boldsymbol{h}}_{rk}^{*} \tilde{ \boldsymbol{h}}_{rb}^{\sf T}  \}$ and $\boldsymbol{R}_{bk}^{d} \defeq \mathbb{E}\{\tilde{ \boldsymbol{h}}_{ab} \tilde{ \boldsymbol{h}}_{ak}^{\sf H}  \}$.
\end{lemma}

%%------------跨栏公式
\newcounter{mytempeqncnt}
\begin{figure*}[!t]
	\normalsize
	\begin{equation}\label{eq:whole}
%		\mathcal{I}\left(\hat{ h}_{a} ; \hat{ h}_{b}\mid \hat{ h}_{e_{k}}\right)
		R_{k}= 
		\log_{2} \left(\frac{\left(P_{\mathrm{B} } g_{u}g_{e}^{k} + (P_{\mathrm{B} } g_{u} + g_{e}^{k}||\boldsymbol{w}||^2)\sigma^2 + ||\boldsymbol{w}||_{2}^2 \sigma^4 -P_{\mathrm{B}} |g_{ue}^{k}|^2  \right) \left(g_{u}g_{e}^{k} + (g_{u}+g_{e}^{k})\sigma^2 + \sigma^4 - |g_{ue}^k|^2   \right)
	}{\left[(P_{\mathrm{B} } + ||\boldsymbol{w}||_{2}^2) \left( g_{u}g_{e}^{k} + (g_{u}+g_{e}^k)\sigma^2 + \sigma^4 - |g_{ue}^{k}|^2\right) -     P_{\mathrm{B} }\sigma^2 (g_{e}^k+ \sigma^2)  \right](g_{e}^{k} + \sigma^2)\sigma^2 }\right).
	\end{equation}
	\hrulefill
\end{figure*}
\begin{IEEEproof}
%	This proof is similar to Lemma~1 in \cite{22Globecom} and omitted 
%	due to the space limitation.
	See Appendix~\ref{sec:covariance_calculation}. 
\end{IEEEproof}

\begin{remark}
	From the KGR expression in (\ref{eq:whole}), we can observe that 
%	the KGR is decided by 
%	the beamforming vectors, 
%	%	i.e. $\bar{\boldsymbol{w}}$ and $\boldsymbol{v}$, 
%	the spatial correlation matrices at Alice and RIS,
%	%	i.e. $\boldsymbol{R}_{S}$, $\bar{\boldsymbol{R}}_{u}$, 
%	and the channel covariance matrices between Bob and Eve.
%	In addition, 
	the KGR is only related to the beamforming vectors and
	statistical channel information, i.e., the covariance matrices. Since the covariance matrices alter slowly in dense scattering environments~\cite{yang2020asymptotic}, we assume that these matrices
	 have been obtained from the previous several time slots by using existing methods, such as the general maximum-likelihood
	 estimation in~\cite{neumann2018covariance}, and we focus on the beamforming optimization to improve the PKG performance.
\end{remark}

\begin{remark}
	If the existing i.i.d. channel model assumptions are adopted to optimize the RIS reflection coefficients, the spatial correlation matrices $\boldsymbol{R}_{S}$ and $\boldsymbol{R}_{I}$ will be the identity matrices. Hence, the $\boldsymbol{w}^{\sf T}\boldsymbol{R}_{S}\boldsymbol{w}^{*}$ and $\boldsymbol{v}^{\sf H}\tilde{\boldsymbol{R}}_{I}\boldsymbol{v}$ in (\ref{eq:whole}) are calculated as $P_{\rm A}$ and $N$, respectively.
%	 and the KGR is decided by the channel covariances between Bob and Eve. 
	 For the case of $\boldsymbol{R}_{bk}^{r}=\rho_{k}\boldsymbol{I}_{N\times N}$ and $\boldsymbol{R}_{bk}^{d} = \rho_{k}\boldsymbol{I}_{M\times M}$, where $\rho_{k}$ is the channel correlation between Eve and Bob, the beamforming design is considered to be independent of the KGR.
%	 and the beamforming $\boldsymbol{w}$ and $\boldsymbol{v}$ cannot be designed. 
	 This observation highlights the importance of taking into account the spatial correlation. 
\end{remark}

Thus, the beamforming design for spatially correlated channel models could be formulated as 
\begin{align}
	\max _{\boldsymbol{w}, \boldsymbol{v}} \
	&\min_{k}\ \left\{\mathcal{I}\left(\hat{ h}_{a} ; \hat{ h}_{b}\mid \hat{ h}_{e_{k}}\right)\right\} \notag \\
	\ \text { s.t. } 
	&\ \text {C1}{:}\,||\boldsymbol{w}||_{2}^2 \leq P_\mathrm{A} ,\notag \\
	&\ \text {C2}{:}\,|v_{n} |\leq 1 , \forall n\in \{1,\cdots N\}, \label{problem:P1}
\end{align}
where $\text {C1}$ indicates the transmit beamforming is constrained by the maximum transmit power budget at the BS, while $\text {C2}$
represents the modulus constraint of each RIS reflection coefficient. 

%\subsection{Problem Transformation}
It could be found that in problem (\ref{problem:P1}), the variables $\boldsymbol{w}$ and  $\boldsymbol{v}$ are in high-order and coupled in the objective function.
To tackle this problem, we first simplify the optimization problem by using the following lemma.
%\begin{align}
%	\mathcal{P}_{1}:\underset{\boldsymbol{w}, \boldsymbol{v}}{\operatorname{max}} \   \ &x - \frac{|y_{k}|^2}{z_{k}+\sigma^2}
%	\label{P1} \\
%	\operatorname{subject\ to}\ &|v_{n} |= 1 , \forall n\in \{1,\cdots N\}, \tag{\ref{P1}{a}} \label{P1-a} \\
%	& ||\boldsymbol{w}||^2 = P_\mathrm{A}, \tag{\ref{P1}{b}} \label{P1-b} 
%\end{align}
\begin{lemma}\label{lemma-transform}
	The original problem (\ref{problem:P1}) is equivalent to the following 
%	max-min fractional optimization 
	problem:
	\begin{align}
		\max _{\bar{\boldsymbol{w}}, \boldsymbol{v}} \
		&\min_{k} \left\{f_{k}(\bar{\boldsymbol{w}},\boldsymbol{v}) \right\}\notag \\
		\quad &\text { s.t. } 
		\text{C1},\text {C2},\label{eq:transform}
	\end{align}
	where $f_{k}(\bar{\boldsymbol{w}},\boldsymbol{v}) = \bar{\boldsymbol{w}}^{\sf H} \boldsymbol{R}_{S} \bar{\boldsymbol{w}} \boldsymbol{v}^{\sf H} \bar{\boldsymbol{R}}_{u} \boldsymbol{v}  -\frac{|\bar{\boldsymbol{w}}^{\sf H} \boldsymbol{R}_{S} \bar{\boldsymbol{w}} \boldsymbol{v}^{\sf H} \bar{\boldsymbol{R}}_{bk}^{r} \boldsymbol{v} +\bar{\boldsymbol{w}}^{\sf H} \bar{\boldsymbol{R}}_{bk}^{d} \bar{\boldsymbol{w}}|^2}{\bar{\boldsymbol{w}}^{\sf H} \boldsymbol{R}_{S} \bar{\boldsymbol{w}} \boldsymbol{v}^{\sf H} \bar{\boldsymbol{R}}_{k} \boldsymbol{v} + \sigma^2}$,
%	\begin{align}
%		f_{k}(\bar{\boldsymbol{w}},\boldsymbol{v}) = \bar{\boldsymbol{w}}^{\sf H} \boldsymbol{R}_{S} \bar{\boldsymbol{w}} \boldsymbol{v}^{\sf H} \bar{\boldsymbol{R}}_{u} \boldsymbol{v}  -\frac{|\bar{\boldsymbol{w}}^{\sf H} \boldsymbol{R}_{S} \bar{\boldsymbol{w}} \boldsymbol{v}^{\sf H} \bar{\boldsymbol{R}}_{bk}^{r} \boldsymbol{v} +\bar{\boldsymbol{w}}^{\sf H} \bar{\boldsymbol{R}}_{bk}^{d} \bar{\boldsymbol{w}}|^2}{\bar{\boldsymbol{w}}^{\sf H} \boldsymbol{R}_{S} \bar{\boldsymbol{w}} \boldsymbol{v}^{\sf H} \bar{\boldsymbol{R}}_{k} \boldsymbol{v} + \sigma^2}, \label{eq:f_k}
%	\end{align}
%	and 
	$\bar{\boldsymbol{R}}_{u} = \beta_{ r }     \boldsymbol{R}_{I} ^{ {\sf T} } \circ   \boldsymbol{R}_{I}           \boldsymbol + \frac{\beta_{ ab }}{N} \boldsymbol{I}$, $\bar{\boldsymbol{R}}_{bk}^{r} = 
	\sqrt{\beta_{ r } \beta_{ r }^{k}} \tilde{\boldsymbol{R}}_{bk}^{r}$, $\bar{\boldsymbol{R}}_{bk}^{d} = \sqrt{\beta_{ ab } \beta_{ ak }}   \tilde{\boldsymbol{R}}_{bk}^{d}$, $\bar{\boldsymbol{R}}_{k} = 	\beta_{ r }^{k}    \boldsymbol{R}_{I} ^{ {\sf T} } \circ   \boldsymbol{R}_{I}           + \frac{\beta_{ ak }}{N} \boldsymbol{I}$, and the variable is defined as $\bar{\boldsymbol{w}} \defeq \boldsymbol{w}^{*}$ for the sake of notational simplicity.
\end{lemma}
\begin{IEEEproof}
	% 	See Appendix~\ref{proof:lemma-transform}.
	It can be proved by using Appendix~B in \cite{22Globecom} and defining $x = f_{k}(\bar{\boldsymbol{w}},\boldsymbol{v})$. 
\end{IEEEproof}

%\begin{remark}
%	From the objective function in (\ref{eq:transform}), we could observe that the KGR is determined by the difference between the two terms, where the first term is decided by the legitimate parties' channels and the second term is decided by Eve's channels. 
%		Also, with the the information leaked to Eve grows, resulting in higher $  \boldsymbol{v}^{\sf H} \bar{\boldsymbol{R}}_{bk}^{r} \boldsymbol{v} $ and $\bar{\boldsymbol{w}}^{\sf H} \bar{\boldsymbol{R}}_{bk}^{d} \bar{\boldsymbol{w}}$, the KGR would decrease. It unveils the importance of optimizing the beamforming vactors $\boldsymbol{w}$ and $\boldsymbol{v}$ to reduce the information leakage.
%\end{remark}
An intuitive solution to tackle problem (\ref{eq:transform}) is to apply the existing ASSG algorithm.
Specifically, the ASSG algorithm can be applied to alternatingly solve for $\boldsymbol{w}$ and $\boldsymbol{v}$ while fixing the other variable, yielding two subproblems with respect to  $\boldsymbol{w}$ and $\boldsymbol{v}$, respectively.
In each subproblem, 
%by defining $\boldsymbol{W}=\boldsymbol{w}\boldsymbol{w}^{\sf H}$ and $\boldsymbol{V}=\boldsymbol{v}\boldsymbol{v}^{\sf H}$, 
SDR-SCA with Gaussian randomization in \cite{21JiTVT} can be then leveraged to convexify the
problem.
However, in the ASSG algorithm, a series of SDP problems generated
by the SCA method need to be solved at each alternating iteration.
In fact, the number of optimization variables in each SDP problem is the square of the number of RIS elements or the BS antennas, contributing to a high-dimensional optimization problem~\cite{2010Semidefinite}. 
At the same time, in RIS-assisted wireless communication systems,
the RIS is often equipped with a large number of elements \cite{pan2021reconfigurable}.
The required computational effort may be unaffordable when ASSG is applied to solve Problem (\ref{eq:transform}), 
%In other words, the ASSG algorithm is not scalable, 
which motivates us to develop a new low-complexity
algorithm for RIS-assisted 
PKG systems.

\section{Impact of Different Beamforming Methods on PKG Performance}
%In Sec. II, we have obtained the optimal solution for reaching the maximum KGR, which is determined by the spatial correlation matrices. 
Before solving Problem  (\ref{eq:transform}), in this section, we aim to compare the PKG performance under the existing assumptions of the i.i.d. channel model~[15]--[17] and the spatially correlated model when Eve experiences independent fading channels.

%We consider a special case where the channel covariance matrices between Bob and Eve's $k$-th antenna are denoted as $\boldsymbol{R}_{bk}^{r} = \rho_{k}\boldsymbol{I}_{N\times N}$ and $\boldsymbol{R}_{bk}^{d} = \rho_{k}\boldsymbol{I}_{M\times M}$, where
%$\rho_{k}$ is the correlation coefficient \cite{2021Sum}. 

%Then, the equivalent form in (\ref{eq:transform}) can be rewritten as 
%\begin{align}
%	f_{k}(\bar{\boldsymbol{w}},\boldsymbol{v}) &= \bar{\boldsymbol{w}}^{\sf H} \boldsymbol{R}_{S} \bar{\boldsymbol{w}} \left(\beta_{ r } \boldsymbol{v}^{\sf H}   \tilde{\boldsymbol{R}}_{I}             \boldsymbol{v} + \beta_{ ba } \right) -\rho_{k}^2\left(\bar{\boldsymbol{w}}^{\sf H} \boldsymbol{R}_{S} \bar{\boldsymbol{w}}\right)^2  
%	 \times   \frac{| \boldsymbol{v}^{\sf H}   \tilde{\boldsymbol{R}}_{I}        \boldsymbol{v}\sqrt{\beta_{ r } \beta_{ r }^{k}} +
%		\sqrt{\beta_{ ba } \beta_{ ka }}
%		|^2  }
%	{(\bar{\boldsymbol{w}}^{\sf H} \boldsymbol{R}_{S} \bar{\boldsymbol{w}})(\beta_{ r }^{k} \boldsymbol{v}^{\sf H}   \hat{\boldsymbol{R}}_{I}           \boldsymbol{v} + \beta_{ ka } ) + \sigma^2} \label{f_k}.
%\end{align}
%
%By analyzing (\ref{eq:transform}), we have the following Lemma.
%\begin{lemma}\label{theorem:analysis}
%	The secret key rate increases monotonically
%	with $\bar{\boldsymbol{w}}^{\sf H} \boldsymbol{R}_{S} \bar{\boldsymbol{w}}$ and $\boldsymbol{v}^{\sf H}   \tilde{\boldsymbol{R}}_{I}             \boldsymbol{v}$, respectively. 
%\end{lemma}
%\begin{IEEEproof}
%	See Appendix~\ref{sec:analysis}.
%\end{IEEEproof}

\begin{lemma} \label{theorem:analysis}
In the case where the eavesdropping channels are independent of the legitimate
channels, the KGR increases monotonically with $ \bar{\boldsymbol{ w }}^{\sf H } \boldsymbol{R}_S \boldsymbol{w}(\beta_{ r }  \boldsymbol{v}^{\sf H} \tilde{\boldsymbol{R}}_{I}\boldsymbol{v} + \beta_{ ab } ) $. 
\end{lemma}
\begin{IEEEproof}
%	See Appendix B.
See Lemma~1 in~\cite{22Globecom}.
\end{IEEEproof}
\begin{remark}\label{remark:opt}
	We first note that the channel correlation between the uplink and downlink channels is $\mathcal{R}_{ab} = \sqrt{P_{\rm B}} \bar{\boldsymbol{ w }}^{\sf H } \boldsymbol{R}_S \boldsymbol{w}(\beta_{ r }  \boldsymbol{v}^{\sf H} \tilde{\boldsymbol{R}}_{I}\boldsymbol{v} + \beta_{ ab } )$. Thus, when optimizing the KGR, 
	the overall channel correlations is increased with the assistance of RIS.
	In addition, since $ \bar{\boldsymbol{ w }}^{\sf H } \boldsymbol{R}_S \boldsymbol{w}$ and $\beta_{ r }  \boldsymbol{v}^{\sf H} \tilde{\boldsymbol{R}}_{I}\boldsymbol{v} + \beta_{ ba }$ are both non-negative, solving Problem (\ref{eq:transform}) is equivalent to maximize these two terms
	separately. 
%	The optimal value 
	Thus, the optimal solution to Problem (\ref{eq:transform}) is $\bar{\boldsymbol{w}}_{\rm opt} = \sqrt{P_{A}}\boldsymbol{u}_{\lambda_{\max }}$ and $v_{n}=e^{j\theta_{n}}$ with $\theta_{n}=\theta,\forall n\in \{1,\cdots,N\}$, where $\boldsymbol{u}_{\lambda_{\max }}$ is the dominant eigenvector of the matrix $\boldsymbol{R}_{S}$ corresponding to its maximum eigenvalue $\lambda_{\max }$ and $\theta$ could take on any value in the interval of $[0,2\pi)$.
\end{remark}

%Based on this observation, we aim to compare the KGR under different channel assumptions. 

%$  g_{u} = ( \boldsymbol{w}^T  \boldsymbol{R}_{S}  \boldsymbol{w}^{ * })(\beta_{ r } \boldsymbol{v}^H   ( \boldsymbol{R}_{I} ^{ T } \circ   \boldsymbol{R}_{I})           \boldsymbol{v} + \beta_{ d } )$
%
%$ g_{ue}^{k} = (\boldsymbol{w}^T  \boldsymbol{R}_{S}  \boldsymbol{w}^{ * })\boldsymbol{v}^H \tilde{\boldsymbol{R}}_{j}^{r}\boldsymbol{v}\sqrt{\beta_{ r } \beta_{ r }^{j}} + \boldsymbol{w}^T  \tilde{\boldsymbol{R}}_{j}^{d}  \boldsymbol{w}^{ * }\sqrt{\beta_{ d } \beta_{ d }^{k}} $
%
%$ g_{e}^{k}= ( \boldsymbol{w}^T  \boldsymbol{R}_{S}  \boldsymbol{w}^{ * })(\beta_{ r }^{k} \boldsymbol{v}^H   ( \boldsymbol{R}_{I} ^{ T } \circ   \boldsymbol{R}_{I})           \boldsymbol{v} + \beta_{ d }^{k} )  $

\subsection{KGR under Different Assumptions of Channel Model at BS}\label{sec:KGR-M}
As shown in Lemma~\ref{theorem:analysis}, the KGR is proportional to $\bar{\boldsymbol{w}}^{\sf H} \boldsymbol{R}_{S} \bar{\boldsymbol{w}}$.  
Under the assumption of the i.i.d. model, the spatial correlation matrix $\boldsymbol{R}_{S}$ is the identity matrix. In this case, the design of transmit beamforming is independent of KGR. As such, random beamforming $\widetilde{\boldsymbol{w}}=\sqrt{P_\mathrm{A}}\widetilde{\boldsymbol{w}}_{0}/||\widetilde{\boldsymbol{w}}_{0}||_{2} $ is applied without loss of generality and optimality, where the entries in  $\widetilde{\boldsymbol{w}}_{0}$ are i.i.d. random variables with zero mean. Then, the expectation of $\widetilde{\boldsymbol{w}}^{\sf H} \boldsymbol{R}_{S} \widetilde{\boldsymbol{w}}$ can be calculated as $\mathbb{E}\{\widetilde{\boldsymbol{w}}^{\sf H} \boldsymbol{R}_{S} \widetilde{\boldsymbol{w}}\} = P_\mathrm{A}$, 
%	\begin{align}
%		\mathbb{E}\{\widetilde{\boldsymbol{w}}^T \boldsymbol{R}_{S} \widetilde{\boldsymbol{w}}^*\} = P_{A},
%	\end{align}
which is independent of the antenna number and the spatial correlation at the BS. 
To investigate the performance loss caused by the design based on the i.i.d. channel assumption, we focus on a typical implementation model of multiple antennas for massive multiple-input multiple-output (MIMO). We consider a general uniform planar array (UPA) model, where the spatial correlation matrix can be approximated as $\boldsymbol{R}_{S} \approx \boldsymbol{R}_{h} \otimes \boldsymbol{R}_{v}$ \cite{choi2014bounds},
%\begin{align}
%	\boldsymbol{R}_{S} \approx \boldsymbol{R}_{h} \otimes \boldsymbol{R}_{v},  \label{RU}
%\end{align} 
where $ \boldsymbol{R}_{h} \succeq \mathbf{0} $ and $ \boldsymbol{R}_{v} \succeq \mathbf{0} $ are the covariance matrices of the horizontal and the vertical uniform linear array (ULA), respectively. 
%Then, by adopting the exponential correlation model that $ [\boldsymbol{R}_{l}]_{n,m} = \rho^{| n - m|} $, we have the following lemma.
%For the ULA spatial correlation matrix, we could model the $ \boldsymbol{R}_{l} $ as an exponential model as a Toeplitz matrix with each element $ \rho^{| i-j|} $, which is the common way to model the ULA spatial correlation matrices. 
The ULA spatial correlation is modeled as a Toeplitz matrix with each element $[\boldsymbol{R}_{l}]_{i,j} = \rho^{| i-j|},l\in \{h,v\} $, where $ 0\leq\rho \leq 1 $ is the correlation index among the antennas. 
Given the optimal transmit beamforming $\bar{\boldsymbol{w}}_{\rm opt}$, we further characterize $\bar{\boldsymbol{w}}_{\rm opt}^{\sf H} \boldsymbol{R}_{S} \bar{\boldsymbol{w}}_{\rm opt}$ via the following lemma. 
\begin{lemma}\label{lemma-analysis}
	For a UPA model, the upper and lower bounds for $\bar{\boldsymbol{w}}_{\rm opt}^{\sf H} \boldsymbol{R}_{S} \bar{\boldsymbol{w}}_{\rm opt}$ are given by 
	\begin{align}
		f_{l}(N_{\mathrm{H}}^{t},N_{\mathrm{V}}^{t},\rho)  \leq  \bar{\boldsymbol{w}}_{\rm opt}^{\sf H} \boldsymbol{R}_{S} \bar{\boldsymbol{w}}_{\rm opt} \leq   f_{u}(N_{\mathrm{H}}^{t},N_{\mathrm{V}}^{t},\rho), 
	\end{align}
	where $ f_{l}(N_{\mathrm{H}}^{t},{N_{\mathrm{V}}^{t}},\rho) = P_\mathrm{A}\frac{ \left( {N_{\mathrm{H}}^{t}}(1 - \rho^2) - 2\rho(1 - \rho^{N_{\mathrm{H}}^{t}})   \right)  }{ {N_{\mathrm{H}}^{t}}{N_{\mathrm{V}}^{t}}(1 - \rho)^4 } \times \left( {N_{\mathrm{V}}^{t}}(1 - \rho^2) - 2\rho(1 - \rho^{N_{\mathrm{V}}^{t}})   \right)$ and $ f_{u}({N_{\mathrm{H}}^{t}},{N_{\mathrm{V}}^{t}},\rho) =P_\mathrm{A} \frac{ (1+\rho^2)(1 - \rho^{{N_{\mathrm{H}}^{t}}-1} )(1 - \rho^{{N_{\mathrm{V}}^{t}} - 1} ) }{ (1 - \rho)^2 } $. $ N_{\mathrm{H}}^{t}$ and $N_{\mathrm{V}}^{t}$ are the number of antennas at horizontal and vertical domains, respectively.
\end{lemma}
\begin{IEEEproof}
	See Lemma 3 in the conference version of this paper \cite{22Globecom}.  %Appendix~\ref{sec:proof-lemma-analysis} 
\end{IEEEproof}
This lemma shows that both the upper and lower bounds increase monotonically with the correlation coefficients $\rho$, the number of antennas $N_{\mathrm{H}}^{t}$, and $N_{\mathrm{V}}^{t}$. This is because the SNR of the combined channel gain increases with the spatial correlation. 
Specifically, when $\rho=0$, the bounds are  
$f_{l}=f_{u}=P_{\rm A}$,
%$f_{l}({N_{\mathrm{H}}^{t}},{N_{\mathrm{V}}^{t}},0)=f_{u}({N_{\mathrm{H}}^{t}},{N_{\mathrm{V}}^{t}},0)=P_{\rm A}$,
 which means the optimal transmit beamforming and random beamforming achieve the same PKG performance in the i.i.d. fading channels.
%Since $f_{l}(K,L,\rho) \geq 1$ and the equality holds in the case of $\rho=0$, $\boldsymbol{w}_{opt}^T \boldsymbol{R}_{S} \boldsymbol{w}_{opt}^* \geq P_{A}$ and achieve larger KGR gain with the increase of correlation and the number of antennas. 
In addition, it can be observed that both the upper and lower bounds converge to $P_\mathrm{A} (\frac{1+\rho}{1-\rho})^2$ as ${N_{\mathrm{H}}^{t}}\rightarrow \infty$ and ${N_{\mathrm{V}}^{t}}\rightarrow \infty$. This means that when the BS is equipped with a large number of antennas, the KGR depends only on the correlations among the antennas of the BS for a given power. 
Also, the KGR increases monotonically with the correlation coefficient $\rho$. The limit of $\bar{\boldsymbol{w}}_{\rm opt}^{\sf H} \boldsymbol{R}_{S} \bar{\boldsymbol{w}}_{\rm opt}$ for $\rho=1$ is $P_{\rm A}N_{\rm H}^{t}N_{\rm V}^{t}$, which is bounded by the transmit power and the antenna numbers at the BS.

\subsection{KGR under Different Assumptions of Channel Model at RIS}\label{sec:KGR-N}
In Lemma~\ref{theorem:analysis}, the KGR is proportional to $\boldsymbol{v}^{\sf H} \tilde{\boldsymbol{R}}_{I}\boldsymbol{v}$. 
Under the assumption of the i.i.d. channel model adopting in existing works, e.g., [15]--[17], the spatial correlation matrix is $\tilde{\boldsymbol{R}}_{I}=\boldsymbol{I}$. By employing random reflection, the expectation of $\widetilde{\boldsymbol{v}}^{\sf H} \tilde{\boldsymbol{R}}_{I} \widetilde{\boldsymbol{v}}$ is $\mathbb{E}\{\widetilde{\boldsymbol{v}}^{\sf H} \tilde{\boldsymbol{R}}_{I} \widetilde{\boldsymbol{v}}\} = N,$
%	\begin{align}
%		\mathbb{E}\{\widetilde{\boldsymbol{v}}^H \tilde{\boldsymbol{R}}_{I} \widetilde{\boldsymbol{v}}\} = N,
%	\end{align}
where each phase in $\widetilde{\boldsymbol{v}}$ can be drawn from the uniform distribution, i.e., $\widetilde{\theta}_{i} \sim U[0,2\pi),i\in \{1,\cdots,N\}$. 

Taking the spatial correlation model into account, the maximum value of $\boldsymbol{v}^{\sf H} \tilde{\boldsymbol{R}}_{I}\boldsymbol{v}$ is $||\boldsymbol{R }_{I}||_{F}^2 $. 
%To characterize the impact of spatial correlation on RIS.
%the physically feasible spatial correlation model at RIS. 
%\begin{lemma}
%	(Proposition 1 in \cite{Rayleigh-fading}) 
	In isotropic scattering environments, the spatial correlation of RIS is expressed as \cite{Rayleigh-fading}
	\begin{align}
		[\boldsymbol{R}_{I}]_{n, m}=\operatorname{sinc} \frac{2\left\|\boldsymbol{u}_{n}-\boldsymbol{u}_{m}\right\|_{2}}{\lambda}, \ \forall n,m \in  \{1,\cdots,N\},\label{eq:R_I}
	\end{align}
	%where the location of the $n$-th element is $ \boldsymbol{u}_{n} = [0,\ i(n)d_{H},\ j(n)d_{V} ]^T$ and $ i(n)= \bmod \left(n-1, N_{\mathrm{H}}\right) $ and $j(n)=\left\lfloor(n-1) / N_{\mathrm{H}}\right\rfloor$ denote horizontal and vertical indices of element $n$, respectively. 
	%the sinc function is $\operatorname{sinc}(x) = \operatorname{sinc}(\pi x)/(\pi x) $.
	where $\left\|\boldsymbol{u}_{n}-\boldsymbol{u}_{m}\right\|_{2}$ denotes the distance between the $n$-th RIS element and the $m$-th RIS element, $\lambda$ is the wavelength. 
%\end{lemma}
%\emph{Remark 1:} 
%Then, the element in  $\tilde{\boldsymbol{R}}_{I} $ is given by $ [\tilde{\boldsymbol{R}}_{I}]_{i,j}  = \left|\operatorname{sinc} \frac{2\left\|\boldsymbol{u}_{n}-\boldsymbol{u}_{m}\right\|}{\lambda} \right|^2$. 
%
Since the sinc function $\operatorname{sinc}(x) = \operatorname{sinc}(\pi x)/(\pi x)$ is monotonically decreasing in interval $[0,1)$, 
the entries in $ \tilde{\boldsymbol{R}}_{I} $ is larger as the inter-element spacing becomes smaller, when the distance between each element pair fulfills $\left\|\boldsymbol{u}_{n}-\boldsymbol{u}_{m}\right\|_{2} \leq \frac{\lambda}{2}$. 
Moreover, the optimal value of $\boldsymbol{v}^{\sf H}\tilde{\boldsymbol{R}}_{I}\boldsymbol{v}$ satisfies $|| \boldsymbol{R}_{I}||_{F}^2 > N$ because the correlation between the elements always exists in practical RIS systems \cite{Rayleigh-fading}. Hence, the KGR performance of the proposed reflective beamforming is better than the counterpart adopting the assumption of the i.i.d. channel model. 

%Also, in Sec. \ref{sec:simulation}, we will discuss whether optimizing $\boldsymbol{w}$ or $\boldsymbol{v}$ leads to a larger KGR.

\subsection{Security Analysis}
In the above analysis, we assume the channels between Bob and Eve are independent. However, in reality Eve may experience correlated channels w.r.t. Bob if Eve's antennas are close to Bob~\cite{2021Sum}. Therefore, we first analyze the information leakage to Eve if the  correlation between eavesdropping channels and legitimate channels is not considered when optimizing $\boldsymbol{w}$ and $\boldsymbol{v}$. To this end, we have the following Lemma.
\begin{lemma}\label{lemma:security}
	Assume that $\rho_{k}$ is the cross channel
	correlation between Bob and Eve~\cite{2021Sum}, i.e., $\boldsymbol{R}_{bk}^{r}=\rho_{k}\boldsymbol{I}_{N\times N}$ and $\boldsymbol{R}_{bk}^{d} = \rho_{k}\boldsymbol{I}_{M\times M}$.
	Given the $\bar{\boldsymbol{w}}_{\rm opt}$ and $\boldsymbol{v}_{\rm opt}$ in Remark~\ref{remark:opt}, the amount of information leaked to Eve increases with the spatial correlation at the BS and the RIS grows.
\end{lemma}
\begin{IEEEproof}
		Given the $\bar{\boldsymbol{w}}_{\rm opt}$ and $\boldsymbol{v}_{\rm opt}$,
		the second term in (\ref{eq:transform}) is denoted as 
		\begin{align}
		f_{k}^{e} =\rho_{k}^2\left(\lambda_{\max }(\boldsymbol{R}_{S})\right)^2  \frac{ \left(\|\boldsymbol{R}_{I}\|_{F}^2\sqrt{\beta_{ r } \beta_{ r }^{k}} +
			\sqrt{\beta_{ ab } \beta_{ ak }}
			\right)^2  }
		{\lambda_{\max }(\boldsymbol{R}_{S})(\beta_{ r }^{k} \|\boldsymbol{R}_{I}\|_{F}^2+ \beta_{ ak } ) + \sigma^2}.	
	\end{align}
Since $\frac{\partial f_{k}^{e}}{\partial(\lambda_{\max }(\boldsymbol{R}_{S}))} \ge 0 $ and $\frac{\partial f_{k}^{e}}{\partial(\|\boldsymbol{R}_{I}\|_{F}^2)} \ge 0 $, the amount of information leaked to Eve is monotonically increasing for $\lambda_{\max }(\boldsymbol{R}_{S})$ and $\|\boldsymbol{R}_{I}\|_{F}^2$. Also, 
%according to Sec. \ref{sec:KGR-M} and Sec. \ref{sec:KGR-N},
 $\lambda_{\max }(\boldsymbol{R}_{S})$ and $\|\boldsymbol{R}_{I}\|_{F}^2$ increase with the spatial correlation at the BS and the RIS, respectively.
	This completes the proof.	
%	See Appendix~\ref{sec:analysis}.
\end{IEEEproof}
Lemma~\ref{lemma:security} states the connections between spatial correlation and security when the correlation between Bob and Eve's channels are ignored. In the following section, we will consider a more general case, i.e., Problem (\ref{eq:transform}), and propose an effective algorithm to tackle it.

\section{BSUM Algorithm For Maximum Key Generation Rate}
In this section, we propose an BSUM-based algorithm to tackle Problem (\ref{eq:transform}).
We decompose the optimization variables, i.e., $\bar{\boldsymbol{w}}$ and $\boldsymbol{v}$, into independent blocks and update the blocks by successively maximizing a
sequence of approximations of the objective function in (\ref{eq:transform}).

\subsection{The BSUM Algorithm for  Problem~(\ref{eq:transform})}
%In this subsection, 

%Specifically, we use the Dinkelbach's transform to solve the non-convex fractional objective function problem. Then, we apply the bock MM method \cite{razaviyayn2013unified} to iteratively optimize the transmit beamforming and reflect beamforming. 

For the simplicity of algorithm design, we first convert Problem (\ref{eq:transform}) from the complex domain to the real
domain\footnote{We transform the problem into the real domain since the modulus operator of complex number in (\ref{eq:transform}) makes the problem intractable.}.
By defining  $\tilde{\boldsymbol{v}}=\left[\Re\{\boldsymbol{v}\}^{\top}, \Im\{\boldsymbol{v}\}^{\top}\right]^{\top} \in \mathbb{R}^{2 N}$
and $\tilde{\boldsymbol{w}}=\left[\Re\{\bar{\boldsymbol{w}}\}^{\top}, \Im\{\bar{\boldsymbol{w}}\}^{\top}\right]^{\top} \in \mathbb{R}^{2 N}$, the optimization problem (\ref{eq:transform}) is equivalent to 
\begin{align}
\max _{\tilde{\boldsymbol{w}}, \tilde{\boldsymbol{v}}} \
	&\min_{k} \left\{\tilde{f}_{k}(\tilde{\boldsymbol{w}},\tilde{\boldsymbol{v}})\right\} \notag \\
	\ \text { s.t. } 
	&\ \overline{\text{C1}}{:}\, \|\tilde{\boldsymbol{w}}\|_{2}^2 \leq P_\mathrm{A}, \notag \\
	&\ \overline{\text{C2}}{:}\,\tilde{v}_{n}^2 + \tilde{v}_{n+N}^2 \leq 1 , \forall n\in \{1,\cdots N\}, \label{eq:real}
\end{align}
where 
%\begin{align}
%	\tilde{f}_{k}(\tilde{\boldsymbol{w}},\tilde{\boldsymbol{v}}) &= \tilde{\boldsymbol{w}}^{\sf T} \tilde{\boldsymbol{R}}_{S} \tilde{\boldsymbol{w}} \tilde{\boldsymbol{v}}^{\sf T} \tilde{\boldsymbol{R}}_{u} \tilde{\boldsymbol{v}}  - \frac{(\tilde{\boldsymbol{w}}^{\sf T} \tilde{\boldsymbol{R}}_{S}\tilde{\boldsymbol{w}} \tilde{\boldsymbol{v}}^{\sf T} \tilde{\boldsymbol{Q}}_{k}^{r} \tilde{\boldsymbol{v}} + \tilde{\boldsymbol{w}}^{\sf T} \tilde{\boldsymbol{Q}}_{k}^{d}\tilde{\boldsymbol{w}})^2}{\tilde{\boldsymbol{w}}^{\sf T} \tilde{\boldsymbol{R}}_{S} \tilde{\boldsymbol{w}} \tilde{\boldsymbol{v}}^{\sf T} \tilde{\boldsymbol{R}}_{k} \tilde{\boldsymbol{v}} + \sigma^2}  - \frac{(\tilde{\boldsymbol{w}}^{\sf T} \tilde{\boldsymbol{R}}_{S}\tilde{\boldsymbol{w}} \tilde{\boldsymbol{v}}^{\sf T} \tilde{\boldsymbol{P}}_{k}^{r} \tilde{\boldsymbol{v}} + \tilde{\boldsymbol{w}}^{\sf T} \tilde{\boldsymbol{P}}_{k}^{d}\tilde{\boldsymbol{w}})^2}{\tilde{\boldsymbol{w}}^{\sf T} \tilde{\boldsymbol{R}}_{S} \tilde{\boldsymbol{w}} \tilde{\boldsymbol{v}}^{\sf T} \tilde{\boldsymbol{R}}_{k} \tilde{\boldsymbol{v}} + \sigma^2}, \label{eq:f_real}
%\end{align}
%\begin{align}
	$\tilde{f}_{k}(\tilde{\boldsymbol{w}},\tilde{\boldsymbol{v}}) = \tilde{\boldsymbol{w}}^{\sf T} \tilde{\boldsymbol{R}}_{S} \tilde{\boldsymbol{w}} \tilde{\boldsymbol{v}}^{\sf T} \tilde{\boldsymbol{R}}_{u} \tilde{\boldsymbol{v}}  - \frac{(\tilde{\boldsymbol{w}}^{\sf T} \tilde{\boldsymbol{R}}_{S}\tilde{\boldsymbol{w}} \tilde{\boldsymbol{v}}^{\sf T} \tilde{\boldsymbol{Q}}_{k}^{r} \tilde{\boldsymbol{v}} + \tilde{\boldsymbol{w}}^{\sf T} \tilde{\boldsymbol{Q}}_{k}^{d}\tilde{\boldsymbol{w}})^2}{\tilde{\boldsymbol{w}}^{\sf T} \tilde{\boldsymbol{R}}_{S} \tilde{\boldsymbol{w}} \tilde{\boldsymbol{v}}^{\sf T} \tilde{\boldsymbol{R}}_{k} \tilde{\boldsymbol{v}} + \sigma^2}  - \frac{(\tilde{\boldsymbol{w}}^{\sf T} \tilde{\boldsymbol{R}}_{S}\tilde{\boldsymbol{w}} \tilde{\boldsymbol{v}}^{\sf T} \tilde{\boldsymbol{P}}_{k}^{r} \tilde{\boldsymbol{v}} + \tilde{\boldsymbol{w}}^{\sf T} \tilde{\boldsymbol{P}}_{k}^{d}\tilde{\boldsymbol{w}})^2}{\tilde{\boldsymbol{w}}^{\sf T} \tilde{\boldsymbol{R}}_{S} \tilde{\boldsymbol{w}} \tilde{\boldsymbol{v}}^{\sf T} \tilde{\boldsymbol{R}}_{k} \tilde{\boldsymbol{v}} + \sigma^2}$,
%\end{align}
%and the real matrices in (\ref{eq:f_real}) are defined as
%\begin{align}
	$\tilde{\boldsymbol{Q}}_{k}^{l}=\left[\begin{array}{ccccccccc}
		\Re\{\frac{\bar{\boldsymbol{R}}_{bk}^{l} + (\bar{\boldsymbol{R}}_{bk}^{l})^{\sf H}}{2} \} & -\Im\{\frac{\bar{\boldsymbol{R}}_{bk}^{l} + (\bar{\boldsymbol{R}}_{bk}^{l})^{\sf H}}{2} \}  \\
		\Im\{\frac{\bar{\boldsymbol{R}}_{bk}^{l} + (\bar{\boldsymbol{R}}_{bk}^{l})^{\sf H}}{2} \} & \Re\{\frac{\bar{\boldsymbol{R}}_{bk}^{l} + (\bar{\boldsymbol{R}}_{bk}^{l})^{\sf H}}{2} \}  
	\end{array}\right] $,
%\end{align}
%\begin{align}
	$\tilde{\boldsymbol{P}}_{k}^{l}=\left[\begin{array}{ccccccccc}
		\Re\{\frac{\bar{\boldsymbol{R}}_{bk}^{l} - (\bar{\boldsymbol{R}}_{bk}^{l})^{\sf H}}{2j} \} & -\Im\{\frac{\bar{\boldsymbol{R}}_{bk}^{l} - (\bar{\boldsymbol{R}}_{bk}^{l})^{\sf H}}{2j} \}  \\
		\Im\{\frac{\bar{\boldsymbol{R}}_{bk}^{l} - (\bar{\boldsymbol{R}}_{bk}^{l})^{\sf H}}{2j} \} & \Re\{\frac{\bar{\boldsymbol{R}}_{bk}^{l} - (\bar{\boldsymbol{R}}_{bk}^{l})^{\sf H}}{2j}\}  
	\end{array}\right], l\in \left\{r,d\right\}$,
%\end{align}
%\begin{align}
%	\tilde{\boldsymbol{P}}_{k}^{d}=\left[\begin{array}{ccccccccc}
%		\Re\{\frac{\bar{\boldsymbol{R}}_{k}^{d} - (\bar{\boldsymbol{R}}_{k}^{d})^{\sf H}}{2j} \} & -\Im\{\frac{\bar{\boldsymbol{R}}_{k}^{d} - (\bar{\boldsymbol{R}}_{k}^{d})^{\sf H}}{2j} \}  \\
%		\Im\{\frac{\bar{\boldsymbol{R}}_{k}^{d} - (\bar{\boldsymbol{R}}_{k}^{d})^{\sf H}}{2j} \} & \Re\{\frac{\bar{\boldsymbol{R}}_{k}^{d} - (\bar{\boldsymbol{R}}_{k}^{d})^{\sf H}}{2j}\}  
%	\end{array}\right], \notag
%\end{align}
and
%\begin{align}
	$\tilde{\A}=\left[\begin{array}{ccccccccc}
		\Re\{\bar{\A} \} & -\Im\{\bar{\A} \}  \\
		\Im\{\bar{\A}\} & \Re\{\bar{\A}\}
	\end{array}\right], \tilde{\A} \in  \left\{\tilde{\boldsymbol{R}}_{S}, \tilde{\boldsymbol{R}}_{u}, \tilde{\boldsymbol{R}}_{k}\right\} $.
Then, since $\tilde{\boldsymbol{w}}$ and $\tilde{\boldsymbol{v}}$ are coupled in Problem (\ref{eq:real}), 
we utilize the BSUM algorithm \cite{razaviyayn2013unified} to decompose them into independent blocks. Specifically, in each iteration, the BSUM algorithm updates a single block of variables by solving an approximate problem of the original problem. If each approximated problem fulfills some conditions as in \cite{MM-TSP}, the sequence of the objective values converges
and first-order optimality holds upon convergence.
Specifically, according to Sec. III-C in \cite{MM-TSP}, given $\tilde{\boldsymbol{w}}^{(i)}$ in the $i$-th  iteration of BSUM, we could construct a lower bound of $\tilde{f}_{k}(\tilde{\boldsymbol{w}}^{(i)},\tilde{\boldsymbol{v}})$ as   
\begin{align}
		\tilde{f}_{k} (\tilde{\boldsymbol{w}}^{(i)},\tilde{\boldsymbol{v}}) 
	& \geq \tilde{f}_{k} (\tilde{\boldsymbol{w}}^{(i)},\tilde{\boldsymbol{v}}^{(i)}) +
	\left( \nabla_{\tilde{\boldsymbol{v}}}  \tilde{f}_{k}(\tilde{\boldsymbol{w}}^{(i)},\tilde{\boldsymbol{v}}^{(i)})\right)^{{\sf T}} \notag \\
	& \quad \times    \left( \tilde{\boldsymbol{v}} - \tilde{\boldsymbol{v}}^{(i)}
	\right)  - \frac{1}{2}(\tilde{\boldsymbol{v}} - \tilde{\boldsymbol{v}}^{(i)})^{\sf T} \M_{k}^{(i)}
	(\tilde{\boldsymbol{v}} - \tilde{\boldsymbol{v}}^{(i)})
	 \notag \\
	& =- \frac{1}{2} \tilde{\boldsymbol{v}}^{{\sf T}} \M_{k}^{(i)} \tilde{\boldsymbol{v}} + 
	\left(\boldsymbol{p}_{k}^{(i)}\right)^{{\sf T}} \tilde{\boldsymbol{v}} + q_{k}^{(i)},
\end{align}
where $\M_{k}^{(i)}=\bar{L}_{k}^{(i)}\boldsymbol{I}_{2N\times 2N}$ that satisfies $\M_{k}^{(i)} \succeq -\nabla_{\tilde{\boldsymbol{v}}}^{2}\tilde{f}_{k} (\tilde{\boldsymbol{w}}^{(i)},\tilde{\boldsymbol{v}})$, and $\bar{L}_{k}^{(i)}$ is calculated as
%求二阶导求二范数
\begin{align}
	\bar{L}_{k}^{(i)}=&\frac{N}{\sigma^2} 
	\left(\frac{4((\tilde{q}_{w}^{(i)}+N\lambda_{\max }(\hat{\boldsymbol{Q}}_{k}^{(i)}))^2+(\tilde{p}_{w}^{(i)}+N\lambda_{\max }(\hat{\boldsymbol{P}}_{k}^{(i)}))^2)}{\sigma^4}
	\right. \notag \\
	&\left. \times
	(m_{w}^{(i)})^2N\lambda_{\max }\left(\tilde{\boldsymbol{R}}_{k} \tilde{\boldsymbol{R}}_{k}\right)+ 2N  \left(\lambda_{\max }\left(\hat{\boldsymbol{Q}}_{k}^{(i)}(\hat{\boldsymbol{Q}}_{k}^{(i)}+(\hat{\boldsymbol{Q}}_{k}^{(i)})^{\sf T})\right) 
	\right.\right. \notag \\
	& \left.\left.
	+ 
	\lambda_{\max }\left(\hat{\boldsymbol{P}}_{k}^{(i)}(\hat{\boldsymbol{P}}_{k}^{(i)}+(\hat{\boldsymbol{P}}_{k}^{(i)})^{\sf T})\right)\right)+\left(\tilde{q}_{w}^{(i)}+N\lambda_{\max }\left(\hat{\boldsymbol{Q}}_{k}^{(i)}\right)\right) \right. \notag \\
	&\left. \times \lambda_{\max }\left(\hat{\boldsymbol{Q}}_{k}^{(i)}+(\hat{\boldsymbol{Q}}_{k}^{(i)})^{\sf T} \right)  +\left(\tilde{p}_{w}^{(i)}+N\lambda_{\max }\left(\hat{\boldsymbol{P}}_{k}^{(i)}\right)\right)
	\right. \notag \\
	&\left. \times 
	\lambda_{\max }\left(\hat{\boldsymbol{P}}_{k}^{(i)}+(\hat{\boldsymbol{P}}_{k}^{(i)})^{\sf T} \right)
	\right),
\end{align}
where  $m_{w}^{(i)} = (\tilde{\boldsymbol{w}}^{(i)})^{\sf T}\tilde{\boldsymbol{R}}_{S}\tilde{\boldsymbol{w}}^{(i)}$, $\tilde{q}_{w}^{(i)} =(\tilde{\boldsymbol{w}}^{(i)})^{\sf T} \tilde{\boldsymbol{Q}}_{k}^{d}\tilde{\boldsymbol{w}}^{(i)}$, $\tilde{p}_{w}^{(i)} =(\tilde{\boldsymbol{w}}^{(i)})^{\sf T} \tilde{\boldsymbol{P}}_{k}^{d}\tilde{\boldsymbol{w}}^{(i)}$, $\hat{\boldsymbol{Q}}_{k}^{(i)} = m_{w}^{(i)} \tilde{\boldsymbol{Q}}_{k}^{r}$, $\hat{\boldsymbol{P}}_{k}^{(i)} = m_{w}^{(i)} \tilde{\boldsymbol{P}}_{k}^{r}$.
In addition, 
%$\boldsymbol{p}_{k}^{(i)} = 2 \left(\lambda^{(i)}_{S}\right)^2 ( \tilde{\boldsymbol{R}}_{u} \tilde{\boldsymbol{v}}^{(i)}(\tilde{\boldsymbol{v}}^{(i)})^{\sf T} \tilde{\boldsymbol{R}}_{k} \tilde{\boldsymbol{v}}^{(i)} +  \tilde{\boldsymbol{R}}_{k} \tilde{\boldsymbol{v}}^{(i)}(\tilde{\boldsymbol{v}}^{(i)})^{\sf T} \tilde{\boldsymbol{R}}_{u} \tilde{\boldsymbol{v}}^{(i)}) 
%+ 2\sigma^2 \lambda^{(i)}_{S}  \tilde{\boldsymbol{R}}_{u} \tilde{\boldsymbol{v}}^{(i)}
%-
%4\left(\lambda^{(i)}_{S}\right)^2 (\tilde{\boldsymbol{v}}^{(i)})^{\sf T} \tilde{\boldsymbol{Q}}_{k}^{r} \tilde{\boldsymbol{v}}^{(i)}  \tilde{\boldsymbol{Q}}_{k}^{r} \tilde{\boldsymbol{v}}^{(i)} - 2\lambda^{(i)}_{S}\tilde{\boldsymbol{Q}}_{k}^{r} \tilde{\boldsymbol{v}}^{(i)}
%- 
%4\left(\lambda^{(i)}_{S}\right)^2 (\tilde{\boldsymbol{v}}^{(i)})^{\sf T} \tilde{\boldsymbol{P}}_{k}^{r} \tilde{\boldsymbol{v}}^{(i)}  \tilde{\boldsymbol{P}}_{k}^{r} \tilde{\boldsymbol{v}}^{(i)} - 2\lambda^{(i)}_{S}\tilde{\boldsymbol{P}}_{k}^{r} \tilde{\boldsymbol{v}}^{(i)}
%-  \eta \lambda^{(i)}_{S} \tilde{\boldsymbol{R}}_{k} \tilde{\boldsymbol{v}}^{(i)}
%+  \bar{L}_{k}^{(i)} \tilde{\boldsymbol{v}}^{(i)}
%$ 
$\boldsymbol{p}_{k}^{(i)} = \left( \nabla_{\tilde{\boldsymbol{v}}}  \tilde{f}_{k}(\tilde{\boldsymbol{w}}^{(i)},\tilde{\boldsymbol{v}}^{(i)})\right)^{{\sf T}}\tilde{\boldsymbol{v}}^{(i)} + \bar{L}_{k}^{(i)} \tilde{\boldsymbol{v}}^{(i)}$, 
and
$q_{k}^{(i)} = \tilde{f}_{k} (\tilde{\boldsymbol{w}}^{(i)},\tilde{\boldsymbol{v}}^{(i)})
-\left( \nabla_{\tilde{\boldsymbol{v}}}  \tilde{f}_{k}(\tilde{\boldsymbol{w}}^{(i)},\tilde{\boldsymbol{v}}^{(i)})\right)^{{\sf T}}  
\tilde{\boldsymbol{v}}^{(i)} - \bar{L}_{k}^{(i)} (\tilde{\boldsymbol{v}}^{(i)})^{{\sf T}} \tilde{\boldsymbol{v}}^{(i)} 
$. As a result, we have
\begin{align}
	\min_{k} \left\{\tilde{f}_{k}(\tilde{\boldsymbol{w}},\tilde{\boldsymbol{v}})\right\} \geq \min_{k} \left\{- \frac{1}{2} \tilde{\boldsymbol{v}}^{{\sf T}} \M_{k}^{(i)} \tilde{\boldsymbol{v}} + 
	(\boldsymbol{p}_{k}^{(i)})^{{\sf T}} \tilde{\boldsymbol{v}} + q_{k}^{(i)}\right\},
\end{align}
and the optimal solution $\tilde{\boldsymbol{v}}^{(i+1)}$ can be obtained by solving the nonsmooth
convex problem 
\begin{align}
	\max _{\tilde{\boldsymbol{v}}} \
	&\min_{k} \left\{- \frac{1}{2} \tilde{\boldsymbol{v}}^{{\sf T}} \M_{k}^{(i)} \tilde{\boldsymbol{v}} + 
	(\boldsymbol{p}_{k}^{(i)})^{{\sf T}} \tilde{\boldsymbol{v}} + q_{k}^{(i)}\right\} \notag \\
	\ \text { s.t. } 
	&\ \overline{{\text{C2}}} .\label{eq:subproblem-v}
\end{align}

Similarly, when the beamforming vector $\tilde{\boldsymbol{v}}^{(i)}$ is fixed in the $i$-th iteration, we could construct the lower bound of $\tilde{f}_{k} (\tilde{\boldsymbol{w}},\tilde{\boldsymbol{v}}^{(i)})$ with respect to $\tilde{\boldsymbol{w}}$ as 
\begin{align}
	\tilde{f}_{k} (\tilde{\boldsymbol{w}},\tilde{\boldsymbol{v}}^{(i)}) 
	& \geq \tilde{f}_{k} (\tilde{\boldsymbol{w}}^{(i)},\tilde{\boldsymbol{v}}^{(i)}) +
	\left( \nabla_{\tilde{\boldsymbol{v}}}  \tilde{f}_{k}(\tilde{\boldsymbol{w}}^{(i)},\tilde{\boldsymbol{v}}^{(i)})\right)^{{\sf T}}    \notag \\
	& \quad \times \left( \tilde{\boldsymbol{w}} - \tilde{\boldsymbol{w}}^{(i)}
	\right)  - \frac{1}{2}(\tilde{\boldsymbol{w}} - \tilde{\boldsymbol{w}}^{(i)})^{\sf T} \mathbf{L}_{k}^{(i)}
	(\tilde{\boldsymbol{w}} - \tilde{\boldsymbol{w}}^{(i)})
	\notag \\
	& =- \frac{1}{2} \tilde{\boldsymbol{w}}^{{\sf T}} \mathbf{L}_{k}^{(i)} \tilde{\boldsymbol{w}} + 
	(\boldsymbol{a}_{k}^{(i)})^{\sf T} \tilde{\boldsymbol{w}} + b_{k}^{(i)},
\end{align}
where 
%$\boldsymbol{a}_{k}^{(i)} = 4\lambda^{(i)}_{S}(\tilde{\boldsymbol{w}}^{(i)})^{\sf T} \tilde{\boldsymbol{R}}_{S}  (\tilde{\boldsymbol{v}}^{(i)})^{\sf T} \tilde{\boldsymbol{R}}_{u} (\tilde{\boldsymbol{v}}^{(i)})(\tilde{\boldsymbol{v}}^{(i)})^{\sf T} \tilde{\boldsymbol{R}}_{k} \tilde{\boldsymbol{v}}^{(i)}
%+ 2\sigma^2  \tilde{\boldsymbol{R}}_{S} \tilde{\boldsymbol{w}}^{(i)} (\tilde{\boldsymbol{v}}^{(i)})^{\sf T} \tilde{\boldsymbol{R}}_{u} \tilde{\boldsymbol{v}}^{(i)}
%-
%2(\tilde{\boldsymbol{w}}^{(i)})^{\sf T} \tilde{\boldsymbol{R}}_{q}\tilde{\boldsymbol{w}}^{(i)}\tilde{\boldsymbol{R}}_{q}\tilde{\boldsymbol{w}} 
%-
%2(\tilde{\boldsymbol{w}}^{(i)})^{\sf T} \tilde{\boldsymbol{R}}_{p}\tilde{\boldsymbol{w}}^{(i)}\tilde{\boldsymbol{R}}_{p}\tilde{\boldsymbol{w}}^{(i)}
% - 2\lambda^{(i)} \tilde{\boldsymbol{R}}_{S} \tilde{\boldsymbol{w}}^{(i)} (\tilde{\boldsymbol{v}}^{(i)})^{\sf T} \tilde{\boldsymbol{R}}_{k} \tilde{\boldsymbol{v}}^{(i)}
% + \tilde{L}_{k}^{(i)}\tilde{\boldsymbol{w}}^{(i)} $ and $\tilde{\boldsymbol{R}}_{q} = \tilde{\boldsymbol{v}}^{\sf T} \tilde{\boldsymbol{Q}}_{k}^{r} \tilde{\boldsymbol{v}} \tilde{\boldsymbol{R}}_{S} + \tilde{\boldsymbol{Q}}_{k}^{d}$. 
% $b_{k} = \tilde{f}_{k}^{(i)} (\tilde{\boldsymbol{w}}^{(t)},\tilde{\boldsymbol{v}}^{(t)}) +  \left( \nabla_{\tilde{\boldsymbol{w}}}  \tilde{f}_{k}^{(i)}\right)^T   
% \tilde{\boldsymbol{w}}^{(t)} - \tau_{w} (\tilde{\boldsymbol{w}}^{(t)})^{\sf T} \tilde{\boldsymbol{w}}^{(t)} $. 
$\mathbf{L}_{k}^{(i)} = \tilde{L}_{k}^{(i)}\I_{2M\times 2M}$ and $\tilde{L}_{k}^{(i)}$ is calculated as
\begin{align}
		\tilde{L}_{k}^{(i)}=&\frac{2P_{\rm A}}{\sigma^2}\left(\lambda_{\max } \left(\bar{\boldsymbol{Q}}_{k}^{(i)}(\bar{\boldsymbol{Q}}_{k}^{(i)}+(\bar{\boldsymbol{Q}}_{k}^{(i)})^{\sf T}) +(\bar{\boldsymbol{Q}}_{k}^{(i)})^{\sf T} \right.\right. \notag \\
		&\left.\left.
		   \times(\bar{\boldsymbol{Q}}_{k}^{(i)}+(\bar{\boldsymbol{Q}}_{k}^{(i)})^{\sf T})\right) 
		+\lambda_{\max }\left(\bar{\boldsymbol{Q}}_{k}^{(i)}\right)\right. \notag \\
		&\left. \times \lambda_{\max }\left(\bar{\boldsymbol{Q}}_{k}^{(i)} + (\bar{\boldsymbol{Q}}_{k}^{(i)})^{\sf T}\right) + (m_{v}^{(i)})^2\frac{4P_{\rm A}^3}{\sigma^4}  \lambda_{\max }\left(\tilde{\boldsymbol{R}}_{S}\tilde{\boldsymbol{R}}_{S}\right)
		\right. \notag \\
		&\left. \times
		(\lambda_{\max }^2\left(\bar{\boldsymbol{Q}}_{k}^{(i)}\right)+\lambda_{\max }^2\left(\bar{\boldsymbol{P}}_{k}^{(i)}\right))+\lambda_{\max }\left(\bar{\boldsymbol{P}}_{k}^{(i)}\right) \right. \notag \\
		&\left. \times \lambda_{\max }\left(\bar{\boldsymbol{P}}_{k}^{(i)} + (\bar{\boldsymbol{P}}_{k}^{(i)})^{\sf T}\right) +\lambda_{\max } \left(\bar{\boldsymbol{Q}}_{k}^{(i)}(\bar{\boldsymbol{Q}}_{k}^{(i)}+(\bar{\boldsymbol{Q}}_{k}^{(i)})^{\sf T}) 
		\right.\right. \notag \\
		&\left. \left. 
		+ (\bar{\boldsymbol{Q}}_{k}^{(i)})^{\sf T}(\bar{\boldsymbol{Q}}_{k}^{(i)}+(\bar{\boldsymbol{Q}}_{k}^{(i)})^{\sf T})\right)
		 \right),
%		 
%		&4\lambda_{\max }\left(\boldsymbol{F}_{q} \boldsymbol{G}_{q}\right) \lambda_{\max }\left(\boldsymbol{F}_{p} \boldsymbol{G}_{p}\right)+2\left(\lambda_{\max }\left(\boldsymbol{F}_{q}\right) \lambda_{\max }\left(\boldsymbol{G}_{q}\right)\right. 
%		\left.+\lambda_{\max }\left(\boldsymbol{F}_{p}\right) \lambda_{\max }\left(\boldsymbol{G}_{p}\right)\right) \notag \\
%		&+2 \lambda_{\max }\left(\tilde{\boldsymbol{Q}}_{k}^{r}\right) \lambda_{\max }\left(\tilde{\boldsymbol{R}}_{S}\right) +\lambda_{\max }\left(\tilde{\boldsymbol{Q}}_{k}^{d}+\left(\tilde{\boldsymbol{Q}}_{k}^{d}\right)^{T}\right)+2 \eta^{(t)} \tilde{\boldsymbol{{R}}}_{S} \lambda_{\max }\left(\tilde{\boldsymbol{{R}}}_{k}\right),
\end{align}
where $\bar{\boldsymbol{Q}}_{k}^{(i)} = \tilde{\boldsymbol{v}}^{(i)} \tilde{\boldsymbol{Q}}_{k}^{r}\tilde{\boldsymbol{v}}^{(i)} \tilde{\boldsymbol{R}}_{S}+ \tilde{\boldsymbol{Q}}_{k}^{d}$, $\bar{\boldsymbol{P}}_{k}^{(i)} = \tilde{\boldsymbol{v}}^{(i)} \tilde{\boldsymbol{P}}_{k}^{r}\tilde{\boldsymbol{v}}^{(i)} \tilde{\boldsymbol{R}}_{S}+ \tilde{\boldsymbol{P}}_{k}^{d}$, $m_{v}^{(i)} = (\tilde{\boldsymbol{v}}^{(i)})^{\sf T} \tilde{\boldsymbol{R}}_{k} \tilde{\boldsymbol{v}}^{(i)}$.
% $\boldsymbol{F}_{q}=\tilde{\boldsymbol{R}}_{s} (\tilde{\boldsymbol{v}}^{(i)})^{\sf T} \boldsymbol{Q}_{k}^{r} \tilde{\boldsymbol{v}}^{(i)}+\boldsymbol{Q}_{k}^{d}$, $\boldsymbol{F}_{p}=\boldsymbol{\tilde{R}}_{S} (\tilde{\boldsymbol{v}}^{(i)})^{\sf T} \boldsymbol{P}_{k}^{r} \tilde{\boldsymbol{v}}^{(i)}+\boldsymbol{P}_{k}^{d}$, $\boldsymbol{G}_{q}=2 (\tilde{\boldsymbol{v}}^{(i)})^{\sf T} \tilde{\boldsymbol{Q}}_{k}^{r} \tilde{\boldsymbol{v}}^{(i)} \tilde{\boldsymbol{R}}_{s}+\boldsymbol{Q}_{k}^{d}+\left(\boldsymbol{Q}_{k}^{d}\right)^{\sf T}$, and $\boldsymbol{G}_{p}=2 (\tilde{\boldsymbol{v}}^{(i)})^{\sf T} \tilde{\boldsymbol{P}}_{k}^{r} \tilde{\boldsymbol{v}}^{(i)} \tilde{\boldsymbol{R}}_{s}+\tilde{\boldsymbol{P}}_{k}^{d}+\left(\tilde{\boldsymbol{P}}_{k}^{d}\right)^{\sf T}$.
Also, 
$\boldsymbol{a}_{k}^{(i)} = \left( \nabla_{\tilde{\boldsymbol{w}}}  \tilde{f}_{k}(\tilde{\boldsymbol{w}}^{(i)},\tilde{\boldsymbol{v}}^{(i)})\right)^{{\sf T}}\tilde{\boldsymbol{w}}^{(i)} + \bar{L}_{k}^{(i)} \tilde{\boldsymbol{w}}^{(i)}$, 
and
$b_{k}^{(i)} = \tilde{f}_{k} (\tilde{\boldsymbol{w}}^{(i)},\tilde{\boldsymbol{v}}^{(i)})
-\left( \nabla_{\tilde{\boldsymbol{w}}}  \tilde{f}_{k}(\tilde{\boldsymbol{w}}^{(i)},\tilde{\boldsymbol{v}}^{(i)})\right)^{{\sf T}}  
\tilde{\boldsymbol{w}}^{(i)} - \bar{L}_{k}^{(i)} (\tilde{\boldsymbol{w}}^{(i)})^{{\sf T}} \tilde{\boldsymbol{w}}^{(i)}$.
Then, we have
%\begin{align}
	$\min_{k} \left\{\bar{f}_{k}(\tilde{\boldsymbol{w}},\tilde{\boldsymbol{v}}^{(i)})\right\} \geq \min_{k}  \left\{- \frac{1}{2} \tilde{\boldsymbol{w}}^{\sf T} \mathbf{L}_{k}^{(i)} \tilde{\boldsymbol{w}} + 
	(\boldsymbol{a}_{k}^{(i)})^{{\sf T}} \tilde{\boldsymbol{w}} + b_{k}^{(i)}\right\}$
%\end{align}
and the convex problem to find the optimal transmit beamforming   $\tilde{\boldsymbol{w}}^{(i+1)}$ can be described as
\begin{align}
	\max _{\tilde{\boldsymbol{w}}} \
	&\min_{k} \left\{- \frac{1}{2} \tilde{\boldsymbol{w}}^{\sf T} \mathbf{L}_{k}^{(i)} \tilde{\boldsymbol{w}} + 
	(\boldsymbol{a}_{k}^{(i)})^{{\sf T}} \tilde{\boldsymbol{w}} + b_{k}^{(i)}\right\}  \notag \\
	\ \text { s.t. } 
	&\ \overline{\text{C1}} . \label{eq:subproblem-w}
\end{align}

\begin{algorithm}[h]
	\caption{The BSUM Algorithm for Problem (\ref{eq:real}).}
	\label{alg:2}
	\begin{algorithmic}[1]
		\Require
		 Threshold $\varepsilon_{0}$ and 
		 covariance matrices $\tilde{\boldsymbol{R}}_{S},\tilde{\boldsymbol{R}}_{u},\tilde{\boldsymbol{Q}}_{k}^{r},\tilde{\boldsymbol{Q}}_{k}^{d},\tilde{\boldsymbol{P}}_{k}^{r},\tilde{\boldsymbol{P}}_{k}^{d},k\in \{1,\cdots,K\}$;
		\State 
		Set: $i = 0$;
%		$\lambda^{(0)} = \min_{k} \tilde{f}_{k}(\tilde{\boldsymbol{w}}^{(0)}, \tilde{\boldsymbol{v}}^{(0)})$;
		\State Initial:  $\tilde{\boldsymbol{v}}^{(0)}$ and $\tilde{\boldsymbol{w}}^{(0)}$;
		
%		\\
%		
		\Repeat  
%		\State $//$ \textit{The outer-loop of the Dinkelbach transformation.}\\
%		\quad \ Set: $i = 0$;
%		\Repeat
%		\State $//$ \textit{The inner-loop of the BMM algorithm.}
		\State Update $\tilde{\boldsymbol{v}}^{(i+1)}$ by solving Problem  (\ref{eq:subproblem-v});
		\State Update $\tilde{\boldsymbol{w}}^{(i+1)}$ by solving Problem (\ref{eq:subproblem-w});
		\State Calculate the objective value of Problem (\ref{eq:real}) as
		\begin{align}
			R^{(i+1)} = \min_{k} \tilde{f}_{k}(\tilde{\boldsymbol{w}}^{(i+1)}, \tilde{\boldsymbol{v}}^{(i+1)});
		\end{align}
		\State $i \leftarrow i+1$;
		\Until{$|R^{(i)} - R^{(i-1)}|\le \varepsilon_{0}$;}
%		\State  $t \leftarrow t+1$;
%		\State Update 
%		\begin{align}
%			$\lambda^{(t)} = \min_{k} \tilde{f}_{k}(\tilde{\boldsymbol{w}}^{(i)}_{(t)}, \tilde{\boldsymbol{v}}^{(i)}_{(t)})$;
%		\end{align}
%		\Until {$|\lambda^{(t)} - \lambda^{(t-1)}|\le \varepsilon_{0}$;}
%		\State 
%		Project $\tilde{v}_{n}$ as $\tilde{v}_{n} \leftarrow \frac{\tilde{v}_{n}}{\sqrt{\tilde{v}_{n}^2 + \tilde{v}_{n+N}^2}} $, $n \in \{1,\cdots, N\}$.
	\end{algorithmic}
\end{algorithm} 
\subsection{Convergence Analysis}
The overall BSUM algorithm
for solving Problem (\ref{eq:real}) is summarized as Algorithm 1.
After 
%applying the BSUM algorithm,
%the Dinkelbach transformation in the outer loop and the BMM algorithm in the inner loop, 
%we 
obtaining the optimized vectors $\tilde{\boldsymbol{v}}_{*}$ and $\tilde{\boldsymbol{w}}_{*}$, we convert them to the complex domain.
%project the $\tilde{\boldsymbol{v}}$ as $\frac{\tilde{v}_{n}}{(\tilde{v}_{n}^2+v_{n+N}^2)^{\frac{1}{2}}},n\in\{1,\cdots,N\}$, such that it satisfies constraint $\overline{\widehat{\text{C2}}}$.
The convergence analysis of the BSUM algorithm is shown as the following lemma.

\begin{lemma}\label{lemma-conver}
%	The convergence property of the proposed DBMM algorithm is comprised of the following two parts. 
%The convergence property of the proposed DBMM algorithm is comprised of the following two parts.
The objective values of Problem (\ref{eq:real}) achieved
by the iteration sequence $\{\tilde{\boldsymbol{v}}^{(i)},\tilde{\boldsymbol{w}}^{(i)}\}_{i=0}^{\infty}$ are non-decreasing and convergence.

%with iterations in the inner loop and the outer loop are both non-decreasing and convergence.
%i.e., $\{\tilde{\boldsymbol{v}}_{(t)}^{(i)},\tilde{\boldsymbol{w}}_{(t)}^{(i)}\}_{i=1}^{\infty}$ and $\{\tilde{\boldsymbol{v}}_{(t)}^{(i)},\tilde{\boldsymbol{w}}_{(t)}^{(i)}\}_{t=1}^{\infty}$,
%In the inner loop (steps 6-11), i.e., the BMM iteration, the objective values of the sequence of iterations  $\{\tilde{\boldsymbol{v}}_{(t)}^{(i)},\tilde{\boldsymbol{w}}_{(t)}^{(i)}\}_{i=1}^{\infty}$ 
%In the outer loop (steps 4-14), i.e., the Dinkelbach's transform, the objective values of the sequence of iterations  $\{\tilde{\boldsymbol{v}}_{(t)}^{(i)},\tilde{\boldsymbol{w}}_{(t)}^{(i)}\}_{t=1}^{\infty}$ are non-increasing and convergence;

\end{lemma}
\begin{IEEEproof}
	See Appendix~\ref{appendix-conver}.
\end{IEEEproof}

\subsection{Computational Complexity}
In the proposed BSUM algorithm, solving the non-smooth convex problems in (\ref{eq:subproblem-v}) and (\ref{eq:subproblem-w}) in each iteration contributes to the most computational cost.
The standard method to solve non-smooth max-min convex problems is to introduce an auxiliary variable $r$ that yields the problem
\begin{align}
	\max _{\tilde{\boldsymbol{v}},r} \
	&\ r \notag \\
	\ \text { s.t. } 
%	&\ \text{C9}{:}\, - \frac{1}{2} \tilde{\boldsymbol{v}}^{{\sf T}} \M_{k}^{(i)} \tilde{\boldsymbol{v}} + 
%	(\boldsymbol{p}_{k}^{(i)})^{{\sf T}} \tilde{\boldsymbol{v}} + q_{k}^{(i)} \geq r, \notag \\
	&\ \overline{{\text{C2}}},\text{C9}{:}\, - \frac{1}{2} \tilde{\boldsymbol{v}}^{{\sf T}} \M_{k}^{(i)} \tilde{\boldsymbol{v}} + 
	(\boldsymbol{p}_{k}^{(i)})^{{\sf T}} \tilde{\boldsymbol{v}} + q_{k}^{(i)} \geq r.\label{eq:problem-smooth}
\end{align}
Problem (\ref{eq:problem-smooth}) is a convex QCQP problem that can be solved by the standard interior-point method~\cite{1502996}. However, the computational complexity of interior-point method to obtain an $\epsilon$-optimal solution is $\mathcal{O}\left(\left(2N^3 + KN^2 \right)\left(\sqrt{N+K} \log \frac{2(N+K)}{\epsilon}\right)\right)$. Thus, 
 the overall computational cost to solve the original problem is still expensive, especially in the case of large $N$, $M$, or $K$. 

Another algorithm 
to solve the problems in (\ref{eq:subproblem-v}) and (\ref{eq:subproblem-w}) is the projected sub-gradient methods~\cite{polyak1987introduction},\cite{shor2012minimization}. Although these methods have low complexity in each iteration, they suffer from a slow convergence rate in general. Specifically, the required iterations to attain an $\epsilon$-optimal solution is no less than $\mathcal{O}\left(\frac{1}{\epsilon^2}\right)$.
 To address these issues, we further propose a computationally efficient algorithm with a fast convergence rate to solve the problems in (\ref{eq:subproblem-v}) and (\ref{eq:subproblem-w}).
 
\section{Mirror-Prox Method for Solving  Problems (\ref{eq:subproblem-v}) and (\ref{eq:subproblem-w})}
In this section, 
%to solve the problems (\ref{eq:subproblem-v}) and (\ref{eq:subproblem-w}) more efficiently, 
we transform the problems in (\ref{eq:subproblem-v}) and (\ref{eq:subproblem-w}) to convex-concave saddle
point problems and apply the Mirror-Prox
method \cite{Mirror-Prox} to solve the resulting problems more efficiently.

\subsection{Mirror-Prox Method for Problems (\ref{eq:subproblem-v}) and (\ref{eq:subproblem-w})}
First, we can transform the non-smooth max-min problem in (\ref{eq:subproblem-v}) into the following equivalent smooth min-max problem by using the primal-dual transformation \cite{21FangTcom}
 \begin{align}
 \min_{\tilde{\boldsymbol{v}}}\ &\max _{\boldsymbol{y}}  \  \psi^{(i)}(\tilde{\boldsymbol{v}},\boldsymbol{y}) \defeq 
 	\left( \bar{\boldsymbol{\tau}}^{(i)}  \|\tilde{\boldsymbol{v}}\|_{2}^2  + 
 	\mathbf{P}^{(i)} \tilde{\boldsymbol{v}}+\mathbf{q}^{(i)}\right)^{\sf T} \boldsymbol{y}
 	\notag \\%%约束
 	\ \text{s.t.} 
% 	&\ \overline{\widehat{\text{C2}}}, \notag \\
 	&\ \overline{{\text{C2}}},\text{C10}{:}\, y_{k} \geq 0,  \sum_{k=1}^{K} y_{k}= 1, \boldsymbol{y} \in \mathbb{R}^{K \times 1}, \label{eq:mirror-v}
 \end{align}
where $\bar{\boldsymbol{\tau}}^{(i)} = \frac{1}{2}[\bar{L}_{1}^{(i)}, \bar{L}_{1}^{(i)} \cdots, \bar{L}_{K}^{(i)} ]^{\sf T} $ and
 the objective function $\psi^{(i)}(\tilde{\boldsymbol{v}}, \boldsymbol{y})$ is convex in $\tilde{\boldsymbol{v}}$ and concave in $\boldsymbol{y}$.
%Then, according to the Sion's minmax equality theorem \cite{sion1958general}, we have
%\begin{align}
%	\min_{\tilde{\boldsymbol{v}}} \max_{\boldsymbol{y}} \psi^{(i)}(\tilde{\boldsymbol{v}},\boldsymbol{y}) =  \max_{\boldsymbol{y}} \min_{\tilde{\boldsymbol{v}}} \psi^{(i)}(\tilde{\boldsymbol{v}},\boldsymbol{y}) ,
%\end{align}
Then, the optimal solution $\tilde{\boldsymbol{v}}_{\rm opt}$ and $\boldsymbol{y}_{\rm opt}$ corresponds to the saddle point of objective function $\psi^{(i)}(\tilde{\boldsymbol{v}},\boldsymbol{y})$~\cite{sion1958general}. 
%In other words, the optimal solution is satisfied with 
%\begin{align}
%	\psi^{(i)}(\tilde{\boldsymbol{v}}_{\rm opt},\boldsymbol{y}) \leq \psi^{(i)}(\tilde{\boldsymbol{v}}_{\rm opt},\boldsymbol{y}_{\rm opt}) \leq \psi^{(i)}(\tilde{\boldsymbol{v}},\boldsymbol{y}_{\rm opt}).  
%\end{align}
By defining $\boldsymbol{z}=[\tilde{\boldsymbol{v}}^{\sf T}, \boldsymbol{y}^{\sf T}]^{\sf T} $ and $\Psi^{(i)}(\boldsymbol{z}) = [\nabla_{\tilde{\boldsymbol{v}}} \psi^{(i)}(\tilde{\boldsymbol{v}})^{\sf T},- \nabla_{\tilde{\boldsymbol{v}}} \psi^{(i)}(\boldsymbol{y})^{\sf T} ]^{\sf T} $,
the problem (\ref{eq:mirror-v}) is equivalent to solve the following variational inequality problem 
\begin{align}
	\text{Find}\quad &\boldsymbol{z}_{\rm opt} \notag \\
	\text{s.t.}\quad &\Psi^{(i)}(\boldsymbol{z}_{\rm opt})^{\sf T}(\boldsymbol{z} - \boldsymbol{z}_{\rm opt}) \geq 0. \label{eq:vari-ineq}
\end{align}
By analyzing the saddle-point operator $\Psi^{(i)}(\boldsymbol{z})$, we have the following Lemma. 
\begin{lemma}\label{lemma-L}
	The operator $\Psi^{(i)}(\boldsymbol{z})$ is monotone and $L_{v}$-Lipschitz
	continuous, where the Lipschitz parameter is $L_{v}=2\sqrt{N} \|\bar{\boldsymbol{\tau}}^{(i)}\|_{2} +  \max_{k}\  \|\p_{k}^{(i)}\|_{2}$.
\end{lemma}
\begin{IEEEproof}
	See Appendix~\ref{appendix-L}
\end{IEEEproof}
%\begin{remark}
%To solve the above Lipschitz continuous
%variational inequality problem (\ref{eq:vari-ineq}), one may apply the generalized projected gradient methods \cite{bubeck2015convex}, since the operator $\Psi^{(i)}(\boldsymbol{z})$ could be seen as the gradient vector. However, the convergence of this method cannot be guaranteed unless $\Psi^{(i)}$ is strongly monotone. Also, the performance of the gradient-based methods is weakened since it could not exploit
%the specific geometry of the problem. The aforementional problems motivates us to find the algorithm
%with superior performance and convergence guarantee.
%\end{remark}
%In \cite{Mirror-Prox}, a fast algorithm
%for solving the variational inequalities of
%the form (30), i.e., the Mirror-Prox algorithm, was proposed. 
Now, we apply the Mirror-Prox method to solve the 
%variational inequalities of
%the form 
Problem
(\ref{eq:vari-ineq}).
The Mirror-Prox method 
%extends the extragradient algorithm to non-Euclidean geometries to 
solves the variational inequality problem optimally with $\mathcal{O}(\frac{1}{\epsilon})$ convergence rate.
We now briefly outline the Mirror-Prox method and the details of this algorithm could be found in \cite{Mirror-Prox}.

%% Algorithm 2
\begin{algorithm}[h]
	\caption{Mirror-Prox Method for Solving the Convex-concave Saddle Point Problem (\ref{eq:subproblem-v}).}
	\label{alg:3}
	\begin{algorithmic}[1]
		\Require Threshold $\epsilon$, stepsize $\alpha=\frac{1}{2L_{v}}$, and operator $\Psi^{(i)}(\cdot)$;
%		\Ensure $\tilde{\boldsymbol{v}}_{\rm opt}$;
%		\State Define $\boldsymbol{z} = [\tilde{\boldsymbol{v}}^{\sf T},\boldsymbol{y}^{\sf T}]^{\sf T}$
		\State Set: $l=0$;
		\State Initial: $\boldsymbol{z}_{0}=[\tilde{\boldsymbol{v}}^{\sf T}_{0},\boldsymbol{y}_{0}^{\sf T}]^{\sf T}$;
		\Repeat 
		\State 
		$\nabla \phi ({\boldsymbol{r}}_{l+1}^{'}) = \nabla \phi({\boldsymbol{z}}_{l})  - \alpha \Psi^{(i)} ( {\boldsymbol{z}}_{l}) $,
		\State 
		${\boldsymbol{r}}_{l+1}^{'} =\nabla \phi^{-1} (\nabla \phi({\boldsymbol{z}}_{l})  - \alpha \Psi^{(i)} ( {\boldsymbol{z}}_{l}))$,
		\State 
		$ {\boldsymbol{r}}_{l+1} =
		\text{arg min}_{\tilde{\boldsymbol{z}}} 
		D_{\phi} ({\boldsymbol{z}},{\boldsymbol{r}}_{l+1}^{'})
		$,
		\State 
		$\nabla \phi ({\boldsymbol{z}}_{l+1}^{'}) = \nabla \phi({\boldsymbol{z}}_{l})  - \alpha \Psi^{(i)} ( {\boldsymbol{r}}_{l+1}) $,
		\State 
		${\boldsymbol{z}}_{l+1}^{'} =\nabla \phi^{-1} (\nabla \phi({\boldsymbol{z}}_{l})  - \alpha \Psi^{(i)} ( {\boldsymbol{r}}_{l+1}))$,
		\State 
		$ {\boldsymbol{z}}_{l+1} =
		\arg \min_{\tilde{\boldsymbol{z}}} 
		D_{\phi} ({\boldsymbol{z}},{\boldsymbol{z}}_{l+1}^{'})
		$;
		\State Set: $l \leftarrow l+1$;
		\Until {$ D({\boldsymbol{z}}_{l}, {\boldsymbol{z}}_{l+1} )\le \varepsilon$;}
		\State Set: ${\boldsymbol{z}}_{\rm opt} \leftarrow \frac{1}{L} \sum_{l=1}^{L} {\boldsymbol{z}}_{l}$.
	\end{algorithmic}
\end{algorithm} 

The Mirror-Prox algorithm is a variant of the mirror descent algorithm~\cite{21FangTcom},\cite{fastfpp}. 
By denoting $\alpha = 1/\left(2L_{v} \right)$, the overall Mirror-Prox algorithm is comprosed of two iterations of Mirror Descent. As shown in Algorithm 2, steps 4-6 correspond to the first mirror descent step, which starts from $\tilde{\boldsymbol{z}}_{l}$ to ${\boldsymbol{r}}_{l+1}$. Then, steps 7-9 follow the similar procedures and start from $\tilde{\boldsymbol{z}}_{l}$ to $\tilde{\boldsymbol{z}}_{l+1}$, using an operator evaluation at $\boldsymbol{r}_{l+1}$. 
In each mirror descent, the projection is done by the Bregman distance $D_{\phi}(\boldsymbol{z}, \boldsymbol{z}^{'})$, which is to monitor
the local geometry of the constraints to improve the algorithm performance \cite{nemirovskij1983problem}.
 The Bregman distance is defined as
\begin{align}
	D_{\phi}(\boldsymbol{z}, \boldsymbol{z}^{'}) = 
	\phi(\boldsymbol{z}) - \phi(\boldsymbol{z}^{'})
	 - \nabla \phi(\boldsymbol{z}^{'})^{\sf T}(\boldsymbol{z} - \boldsymbol{z}^{'}), \label{eq:Breg}
\end{align}
where $\phi(\boldsymbol{z})$ is the mapping function. According to the structure of $\overline{{\text{C2}}}$ and $\text{C10}$, we select the mapping function as 
%\begin{align}
	$\phi(\boldsymbol{z}) = \frac{1}{2}\|\tilde{\boldsymbol{v}} \|^2
			+ \sum_{k=1}^{K} y_{k}\ln y_{k}$~ \cite{ibrahim2018mirror},
%\end{align}
where the first term is the mirror map for the Euclidean space $\overline{{\text{C2}}}$, and the second term denotes the mirror map for the simplex space $\text{C10}$. 
Thus the $\nabla \phi(\boldsymbol{z})$ and $\nabla^{-1} \phi(\boldsymbol{z})$ is expressed as
	$\nabla \phi(\boldsymbol{z})=\left[\tilde{\boldsymbol{v}},
	\ln{y}_{1}+1,
	\cdots,
	\ln {y}_{K}+1\right]^{\sf T}$
	, and $
	\nabla^{-1} \phi(\tilde{\boldsymbol{z}})=
	\left[\tilde{\boldsymbol{v}},
	\exp \left({y}_{1}-1\right),
	\cdots,
	\exp \left({y}_{K}-1\right)\right]^{\sf T}$.
%\begin{align}
%	\nabla \phi(\boldsymbol{z})&=\left[\tilde{\boldsymbol{v}},
%	\log {y}_{1}+1,
%	\cdots,
%	\log {y}_{K}+1\right]^{\sf T}
%	, \\
%	\nabla^{-1} \phi(\tilde{\boldsymbol{z}})&=
%	\left[\tilde{\boldsymbol{v}},
%	\exp \left({y}_{1}-1\right),
%	\cdots,
%	\exp \left({y}_{K}-1\right)\right]^{\sf T},
%\end{align}
Hence, the Bregman distance in (\ref{eq:Breg}) can be expressed as
\begin{align}
		D_{\phi}(\boldsymbol{z}, \boldsymbol{z}^{'}) = \frac{1}{2}\|\tilde{\boldsymbol{v}} - \tilde{\boldsymbol{v}}^{'} \|^2
		+ \sum_{k=1}^{K} y_{k}\ln\frac{y_{k}}{y_{k}^{'}} - 
		\sum_{k=1}^{K}(y_{k} - y_{k}^{'}). \label{eq:D}
\end{align}
Based on (\ref{eq:D}),
%by using the Bregman distance (\ref{eq:D}), 
the non-Euclidean projection problem in step 6
% can be written as
%\begin{align}
%	 \min_{\tilde{\boldsymbol{v}},\boldsymbol{y}}\  &\frac{1}{2}\|\tilde{\boldsymbol{v}} - \tilde{\boldsymbol{v}}^{'} \|^2
%	+ \sum_{k=1}^{K} y_{k}\log_{2}\frac{y_{k}}{y_{k}^{'}} - 
%	\sum_{k=1}^{K}(y_{k} - y_{k}^{'}) \notag \\
%	\text{ s.t. }\ &\text{C2},\text{C3},\label{eq:proj}
%\end{align}
%which 
can be solved by minimizing the first term in (\ref{eq:D}) with respect to $\tilde{\boldsymbol{v}}$ and the others terms in (\ref{eq:D}) with respect to $\boldsymbol{y}$ separately. 
%\begin{align}
%	\min_{\tilde{\boldsymbol{v}}} &\ \|\tilde{\boldsymbol{v}} - \tilde{\boldsymbol{v}}^{'} \|_{2}, \label{eq:52}\\
%	\min_{\boldsymbol{y}}	&\ \sum_{k=1}^{K} y_{k}\log_{2}\frac{y_{k}}{y_{k}^{'}} - 
%	\sum_{k=1}^{K}(y_{k} - y_{k}^{'}).\label{eq:53}
%\end{align}
%The first problem is an projection to Euclidean space and 
Specifically, the closed-form solution of $\tilde{\boldsymbol{v}}$ to the problem in step 6 is expressed as
\begin{align}
	\tilde{v}_{i}=\left\{\begin{array}{cl}
		\frac{\tilde{v}^{\prime}_{i}}{\left(\tilde({v}^{\prime}_{i})^2 + (\tilde{v}^{\prime}_{N+i})^2\right)^{\frac{1}{2}}}, & u_{i}^{2}+u_{N+i}^{2} \geq 1, \\
		\tilde{v}^{\prime}_{i}, & \text { otherwise, }
	\end{array}   \right.
\end{align}
and 
%the second problem 
%is a
%non-Euclidean projection onto the
% $K$-dimensional probability simplex, and the closed form 
 the optimal solution of $\boldsymbol{y}$  is $\boldsymbol{y}_{\rm opt} = \frac{\boldsymbol{y}^{\prime}}{\|\boldsymbol{y}^{\prime}\|_{1}}$~\cite{bubeck2015convex}.
%\begin{align}
%	\boldsymbol{y}_{\rm opt} = \frac{\boldsymbol{y}^{\prime}}{\|\boldsymbol{y}^{\prime}\|_{1}}.
%\end{align}
The above implementation details can be applied to steps 7-9 in Algorithm~2 directly.

%\subsection{Mirror-Prox Method for Problem (\ref{eq:subproblem-w})}
Similarly to the above steps that tackle the problem in (\ref{eq:subproblem-v}), we can apply Algorithm 2 to tackle the problem in (\ref{eq:subproblem-w}) by replacing $\M_{k}^{(i)}$, $\boldsymbol{p}_{k}^{(i)}$, and $q_{k}^{(i)}$ with $\mathbf{L}_{k}^{(i)}$, $\boldsymbol{a}_{k}^{(i)}$, and $b_{k}^{(i)}$, respectively.

\subsection{Computational Complexity}
For Algorithm 2, the mirror projections are all closed-form operations. The computation cost to solve problem (\ref{eq:subproblem-v}) is dominated by the computation of $\Psi^{(i)}(\cdot)$, where the complexity is $\mathcal{O}(NK)$. Also, the complexity to tackle problem (\ref{eq:subproblem-w}) is $\mathcal{O}(MK)$.
In Sec. VII, we will present the specific running time comparison to verify the effectiveness of our proposed algorithm.

\section{Simulation Results}\label{sec:simulation}
In this section, we provide simulation results to illustrate the PKG performance of the proposed method and the impact of spatial correlation on KGR.

\subsection{Simulation Settings}
%\begin{figure}
%	\centering
%	\includegraphics[width=3.1in]{fig/simulation_model2}
%	\caption{\blue{Simulation setup.}}
%	\label{simulation model}
%\end{figure}
%As shown in Fig. (\ref{simulation model}),
 We consider a three-dimensional coordinate  system where the central point of Alice, Bob, and RIS are located at (5\ m, 0\ m, 20\ m), (3\ m, 100\ m, 0\ m), and (0\ m, 60\ m, 2\ m), respectively~\cite{21JiTVT}.
% \cite{Globecom}.
 We assume that Alice is equipped with a UPA located in $x-z$ plane, and the RIS is equipped with a uniform rectangular array (URA) located in $y-z$ plane.
  The UPA at the BS has $N_{\mathrm{H}}^t=5 $ antennas per row and $ N_{\mathrm{V}}^t=M/5$ antennas per column. 
  The RIS has $N_{\mathrm{H}}^r=5 $ elements per row and $ N_{\mathrm{V}}^r=N/5$ elements per column. 
   The antennas of Eve are randomly distributed within a circle of radius $R$ centered at Bob and the channel correlation coefficient is $\rho_{k}=[J_{0}(2\pi d/\lambda)]^2$, where $d$ is the distance and $J_{0}$ is the first-kind of Bessel function~\cite{21JiTVT}. 
%The distance-dependent large-scale fading is modeled as $T_{0}(d/d_{0})^{-\alpha}$, where $T_{0}$ is the path loss at the reference distance 1\ m. 
The channel between Alice and Bob is generated by (\ref{eq3}) and the large-scale path loss $ \beta_{ba} =\sqrt{\zeta_{0} d_{ba}^{-\alpha_{ba}} } $, where $d_{ba}$, $\zeta_{0}$, and $\alpha_{ba}$ are the distance, path loss at 1 m, and the path loss exponent, respectively. 
The simulation settings are $\alpha_{ba}= \alpha_{e_{k}a}= 4 $, $ \alpha_{ar} = 3.5 $, $\alpha_{br}=\alpha_{e_{k}r} =2$, $\zeta_{0}=-30$ dB, $\sigma^2 = -80$\ dBm, and $\epsilon = 10^{-4}$~\cite{21JiTVT}.
% We perform $1000$ random channel realizations and show the average results in the following subsections.
The results in the following subsections were averaged over $1000$ random channel realizations.

%\subsection{Baseline Schemes}
%To show the effectiveness of the proposed algorithm, we give the baseline schemes as follows. 
%\begin{itemize}
%	\item $\textbf{ASSG algrithm, spatially correlated models}$: This method assumes the channels at both Alice and RIS are both i.i.d. fading \cite{21JiTVT},\cite{22Sum-RIS}. Hence, the covariance matrix $\boldsymbol{R}_{I}$ and $\boldsymbol{R}_{S}$ are both identity matrices. 
%	Based on this model, we jointly optimize the transmit beamforming $\boldsymbol{w}$ and the reflect beamforming $\boldsymbol{v}$ via our proposed algorithm. 
%	
%	\item $\textbf{ASSG algorithm, i.i.d. assumption at RIS}$: This method assumes the channels at both Alice and RIS are both i.i.d. fading \cite{21JiTVT},\cite{22Sum-RIS}. Hence, the covariance matrix $\boldsymbol{R}_{I}$ and $\boldsymbol{R}_{S}$ are both identity matrices. 
%	Based on this model, we jointly optimize the transmit beamforming $\boldsymbol{w}$ and the reflective beamforming $\boldsymbol{v}$ via our proposed algorithm. 
%	
%	\item $\textbf{ASSG algorithm, i.i.d. assumption at RIS}$: This method assumes the channels at both Alice and RIS are both i.i.d. fading \cite{21JiTVT},\cite{22Sum-RIS}. Hence, the covariance matrix $\boldsymbol{R}_{I}$ and $\boldsymbol{R}_{S}$ are both identity matrices. 
%	Based on this model, we jointly optimize the transmit beamforming $\boldsymbol{w}$ and the reflect beamforming $\boldsymbol{v}$ via our proposed algorithm. 
%	
%	
%	
%
%\end{itemize}

\subsection{Convergence of the Proposed Algorithms}
\begin{figure}
	\centering
%	{\includegraphics[width=0.5\textwidth]{fig/converg/Conver_New.eps}}
	{\includegraphics[width=0.45\textwidth]{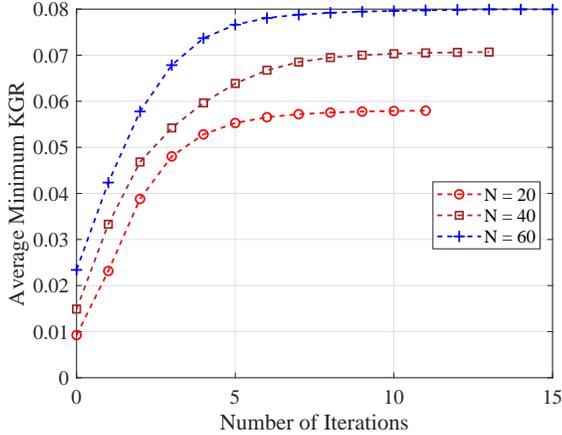}}%F_converg_comp_update1
	\caption{Convergence behavior of the BSUM algorithm under different $N$ when $P_{\rm A}=P_{\rm B}=20$ dBm, $M=15$, $\rho=0.3$, the spacing of two neighbouring RIS elements is $\lambda/4$, and $K=10$.}\label{fig:converg}
\end{figure}

First, we evaluate the convergence behavior of the proposed BSUM with Mirror-Prox algorithm. Fig. \ref{fig:converg} depicts the average minimum KGR versus the number of iterations for various  RIS elements, i.e., for $N=20$, $40$, and $60$. 
It can be observed that having more RIS elements leads to
a slightly slower convergence speed.
As more optimization variables are involved, the more iterations
are required for convergence due to the enlarged solution space.
However, for different values of $N$, the proposed algorithm converges within $15$ iterations on average, which illustrates the practicality of the proposed algorithm.
%
%under different transmit power, the proposed DBMM algorithm monotonically decreases over the iterations and converges within 25 iterations. 
%This observation confirms the complexity efficient of the proposed algorithm, since the computational complexity in each iteration is relatively lower. 
%In addition, the number
%of iterations required for the convergence of the proposed method increases with the transmit power $P_{\rm A}$, as the feasible set of optimization variable $\boldsymbol{w}$ increases. 
%Also, the convergence value
% increases with transmit power $P_{\rm A}$ since the SNR is improved. In addition, it is noted that the KGRs of the proposed DBMM algorithm before and after projecting the transmit beamforming $\boldsymbol{v}$ to $\mathcal{V}$ are almost the same. This is because in the simulation the optimized variable $\boldsymbol{v}$ of our proposed method is always satisfied with the unit modulus constraint. \blue{$M_{\rm H}=5$}

\subsection{KGR versus the Transmit Power}
\begin{figure}
	\centering
	{\includegraphics[width=0.45\textwidth]{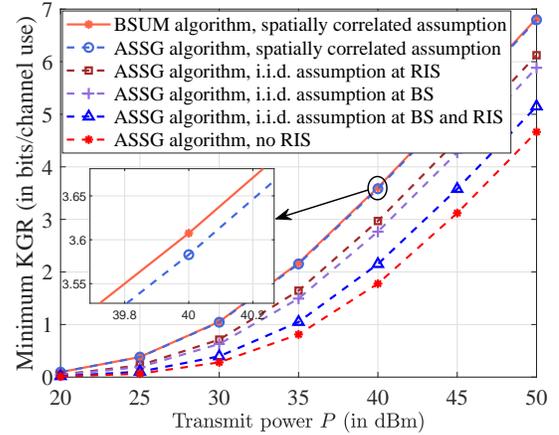}}%power_KGR_dbmm1 Pab/KGR_Pab
	\caption{Average minimum KGR achieved by different algorithms with different channel assumptions when $P_{\rm A}=P_{\rm B}=P$, $N=60$, $M=15$, $K=5$, $\rho=0.2$, and the spacing of two neighbouring RIS elements is $\lambda/4$.}\label{fig:power_KGR_Eve}
\end{figure}
In Fig. \ref{fig:power_KGR_Eve}, the average minimum KGR versus the transmit power under different algorithms and channel assumptions is plotted. First, it can be observed that the KGR at all the settings increases with
the transmit power, since the negative impacts of noises in the channel estimation and key generation are
reduced.
For comparison, some benchmarks are provided: (1) the ASSG algorithm 
based on different channel models; (2) the case without RIS. 
It is noted that the proposed design outperforms these benchmarks.
In particular, the BSUM with Mirror-Prox and ASSG algorithms under correlated channel model assumption lead to a higher KGR than the benchmarks under i.i.d. channel model assumptions. 
Specifically, when $P\geq 30$ dBm, the proposed setting achieves about $5$ dB and $6$ dB transmit power gain compared to the beamforming scheme under the i.i.d.
channel assumption and the optimal transmit beamforming
without RIS, respectively.
%when $P=30$ dBm, the proposed setting achieves about $4$ dB and $5$ dB transmit power
%gain compared to the beamforming scheme 
%under the i.i.d. channel assumption at the RIS and i.i.d. channel assumption at the BS, respectively. 
This is because when correlations
exist between the BS antennas and the RIS elements, the i.i.d.
model fails in capturing this important characteristic which degrades the KGR performance. In contrast, the proposed scheme can effectively exploit the properties of the channels to perform a more precise beamforming. 
It can also be observed that the beamforming design under the i.i.d. assumption at the RIS achieves a higher KGR gain than that under the i.i.d. assumption at the BS. This is because when $\rho=0.2$ and the neighbouring RIS element spacing is $\lambda/4$, having the optimal $\boldsymbol{w}$ is more effective than that of $\boldsymbol{v}$ in combating the noises in $R_{k}$. 
%and this aligns with the analysis in (\ref{eq:transform}).
Finally, we observe that in spatially correlated channels, the BSUM with Mirror-Prox algorithm and ASSG algorithm achieve almost the same KGR under different transmit powers, since both algorithms guarantee to converge to a stationary point. 
However, Gaussian randomization is applied in ASSG algorithm to obtain a rank-1 solution,
 which results in slight KGR degradation on average. 
%These results clearly show the effectiveness of the proposed algorithm in jointly optimizing the transmit beamforming and the reflect beamforming. 

% 为什么优化w更有效？

\subsection{Computation Time Comparision}
\begin{figure}
	\centering
	{\includegraphics[width=0.45\textwidth]{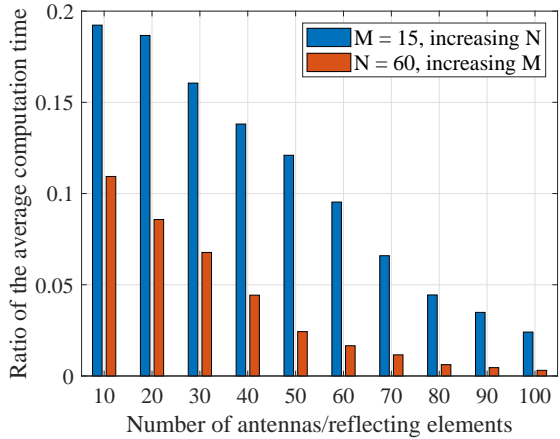}}
	\caption{Ratio of the average computation time of the BSUM algorithm to the ASSG algorithm when $P_{\rm A} = P_{\rm B} = 20$ dBm, $K=4$, the spacing of two neighbouring RIS elements is $\lambda/2$, and $\rho=0.3$.}\label{fig:time}
\end{figure}
%\begin{table}[ht]\label{Table:time}
%	\centering
%	\caption{Ratio of the average computation time of the BSUM algorithm to the ASSG algorithm.}
%	\label{tab:rates}
%	\begin{tabular}{@{}lcccccccccc@{}}
%		\toprule
%		%& \multicolumn{4}{c}{Country List} \\
%		\textbf{Number of antennas/elements}   
%		&                    \textbf{$10$}    
%		& \textbf{$20$} 
%		& \textbf{$30$} 
%		& 
%		\textbf{$40$}
%		& 
%		\textbf{$50$}
%		& 
%		\textbf{$60$}
%		& 
%		\textbf{$70$}
%		& 
%		\textbf{$80$}
%		& 
%		\textbf{$90$}
%		& 
%		\textbf{$100$}
%		\\
%		\toprule
%		\multirow{1}{*}{\textbf{Ratio ($N=60$, increasing $M$)}}
%		&22.4\%
%		&21.0\%
%		&13.4\%
%		&8.6\%
%		&6.7\%
%		&5.6\%
%		&4.3\%
%		&3.2\%
%		&2.7\%
%		&1.9\%
%		\\
%		\midrule 
%		\multirow{1}{*}{\textbf{Ratio ($M=15$, increasing $N$)}}              
%		&13.4\%
%		&11.8\%
%		&11.5\%
%		&11.3\%
%		&11.3\%
%		&10.2\%
%		&10.1\%
%		&9.6\%
%		&7.7\%
%		&7.2\% 
%		\\
%		%BDR, $k=94$                   & 0.218   & 0.120  & 0.082  &  0.065 %\\
%		%		& Without RIS            & 0.481   & 0.492  & 0.491  & 0.488  \\
%		%BDR, $k=94$                   & 0.499   & 0.489  & 0.486  & 0.495\\
%		\bottomrule 
%	\end{tabular}
%\end{table}
In Fig. \ref{fig:time}, we present the ratio of the average computation time of the  BSUM algorithm to that of the ASSG algorithm. As can be observed, the BSUM with Mirror-Prox algorithm consumes much shorter computation time than the ASSG algorithm and
%,
%which indicates the proposed  algorithm significantly outperforms the ASSG algorithm in terms of computation time.
%Also, 
the ratio decreases as the number of the BS antennas or the RIS elements increases.
This is because the dimension of the optimization variable is the square of the BS antennas, i.e., $M^{2}$ or $N^{2}$, in the ASSG algorithm, while the BSUM algorithm optimizes the $M$ or $N$-dimensional vector directly.
%In particular, the BSUM algorithm only consumes $1.9\%$ of the computation time compared to the ASSG algorithm when $ N=60$ and $M=100$.
%In addition, the computation time of the DBMM algorithm and the ASSG algorithm versus the number of RIS elements is plotted in Fig. \ref{fig:Time_N}. It is observed that the DBMM algorithm has
%a superior performance in terms of the computation time as the number of RIS elements increases. 
%Specifically, 
%%compared to the ASSG algorithm, 
%%the proposed algorithm only consumes about 
%the reduction ratios are roughly $90\%$ and $93\%$
%when $N=70$ and $N=100$, respectively.
The above results verify the computational effectiveness of the proposed algorithm. 
%in solving large-scale optimization problems. 

%\begin{figure}
%	\centering
%	{\includegraphics[width=0.55\textwidth]{fig/time/N.eps}}
%	\caption{Computation time versus the number of elements at RIS for different algorithm when $P_{\rm A} = P_{\rm B}=30$ dBm, $M=15$, $K=10$, RIS elements spacing is $\lambda/4$, and $\rho=0.3$.}\label{fig:Time_N}
%\end{figure}

\subsection{The Impact of RIS Elements Number and Size}
\begin{figure}
	\centering
	{\includegraphics[width=0.45\textwidth]{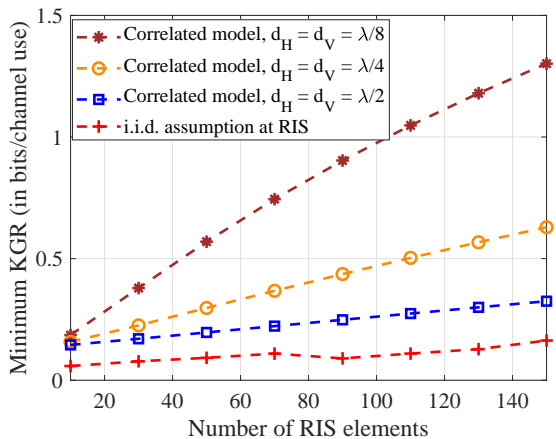}} %KGR_N2
	\caption{Average KGR achieved for different $N$ when $P_{\rm A}=30$ dBm, $P_{\rm B}=20$ dBm, $M=15$, $K=5$, and $\rho=0.2$. }\label{fig:KGR_N}
\end{figure}
Fig. \ref{fig:KGR_N} shows the KGR of different RIS neighbour elements spacing versus the number of RIS elements adopting the proposed algorithm. As can be observed, the KGR of all of these cases increases with the number of RIS elements. This is because with more RIS elements in place, the proposed
design becomes more flexible to create pencil-like energy
focusing beams at the RIS to realize better KGR performance.
In addition, it is noted that with the RIS elements spacing becoming smaller, the minimum KGR increases significantly. The reason behind this is that with smaller elements spacing, the values of the spatial correlation $\boldsymbol{R}_{I}$ are larger, e.g., (\ref{eq:R_I}), contributing to a higher KGR. Finally, it is found that even with $\lambda/2$ RIS element spacing, the KGR of the proposed method is still slightly superior than that adopting the i.i.d. channel assumption. In fact, the correlation among the RIS elements is weak in $\lambda/2$ spacing, although it always exists if $N_{\rm H}^{r} >1$ or $N_{\rm V}^{r} >1$, which can be exploited by the proposed method.

\subsection{The Impact of BS Antennas Number and Correlation} 
\begin{figure}
	\centering
	{\includegraphics[width=0.45\textwidth]{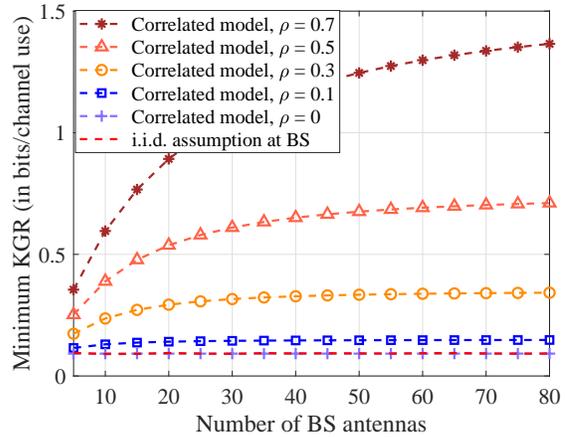}} %KGR_M1
	\caption{Average KGR versus different $M$ when $P_{\rm A}=30$ dBm, $P_{\rm B}=20$ dBm, $K=5$, and the spacing of two neighbouring RIS elements is $\lambda/2$.}\label{fig:KGR_M}
\end{figure}
Fig. \ref{fig:KGR_M} shows the KGR versus the number of the antennas at the BS. The beamforming vectors are optimized by the proposed BSUM with Mirror-Prox algorithm. As can be observed, the KGR of the design method based on the i.i.d. fading model is identical to that of the proposed design when $\rho = 0$, which is independent of the antenna numbers at the BS. For the cases of $\rho>0$, the proposed method can always achieve higher KGR gain, since the upper and lower bounds of the KGR both increase with the spatial correlation between antennas. 
Moreover, with the number of antennas increases, the KGR increases with diminishing returns. This is due to channel hardening~\cite{Rayleigh-fading} and the limited transmit power at the BS. 
%by the maximum transmit power constraint  
Indeed, the greater correlation coefficient $\rho$ contributes to higher converged value, which is consistent with Lemma~\ref{lemma-analysis}.

\subsection{The Impact of Eve Antennas Number and Location} 
\begin{figure}
	\centering
	{\includegraphics[width=0.45\textwidth]{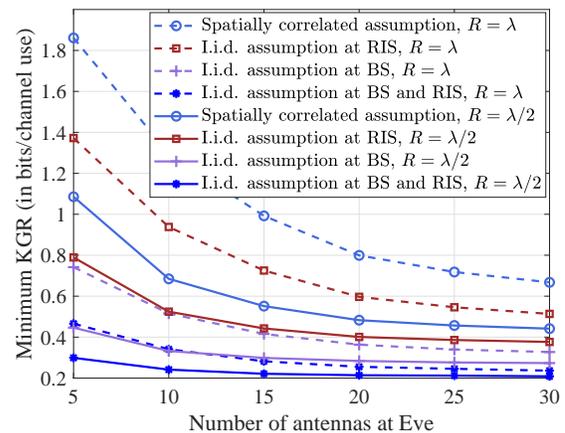}} %K/KGR_K
	\caption{Comparison of the average minimum KGR for different channel assumptions versus the Eve's distribution radius $R$ when $P_{\rm A}=P_{\rm B}=30$ dBm, $N=60$, $M=20$, $\rho = 0.4$, and the spacing of two neighbouring RIS elements is $\lambda/4$.}\label{fig:KGR_K}
\end{figure}

In Fig. \ref{fig:KGR_K}, the average minimum KGR versus the number and the location of the eavesdropper's antennas is plotted. 
%As can be observed, the number and the location of the Eve's antenna has an significant impact on the KGR. 
First, the KGR
decreases when there are more antennas for Eve. This is due to the fact that more antennas at Eve create better quality in the eavesdropping channels. 
%Specifically, given the location radius $R$, Eve's antenns tend to be closer to Bob when it is equipped with 
In addition, Fig. \ref{fig:KGR_K} also reports that the radius of Eve's antennas location, $R$, has a significant impact on the KGR. With a larger radius of Eve's distribution, the average minimum correlation between Eve's antennas and Bob decreases, that contributes to higher KGR. 
Finally, as expected, 
the KGR of the proposed algorithm outperforms those benchmarks that adopt the i.i.d. channel assumption at the BS or the RIS. 
The results above verify the effectiveness of the proposed method
under different eavesdropping conditions.

\subsection{BDR Comparison and Randomness Evaluation} 
\begin{figure}
	\centering
	{\includegraphics[width=0.45\textwidth]{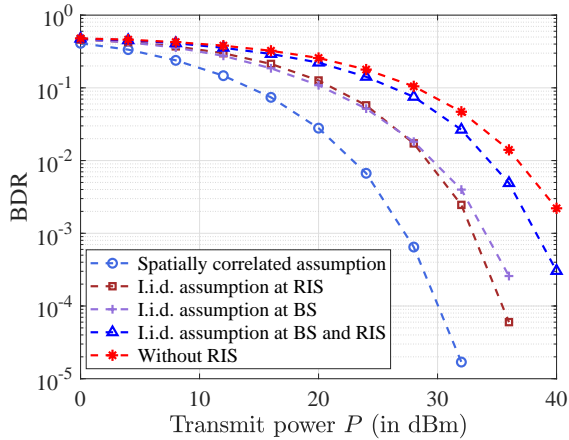}} %BDR/BDR
	\caption{Comparison of the average BDRs for different channel assumptions versus the transmit power $P$.}\label{fig:BDR}
\end{figure}
After the channel probing stage, Alice and Bob quantize their combined channel gains into raw key bits. 
%in the quantization stage. 
The bit disagreement ratio (BDR) denotes the ratio of the number of disagreements bits to the number of total quantized bits. When the BDR becomes higher, the legitimate ends have to consume larger signaling overhead to correct these inconsistent bits.
Fig. \ref{fig:BDR} presents the BDR performance under different channel assumptions using the proposed algorithm and the 1-bit Cumulative Distribution Function (CDF)-based  quantization method~\cite{CDF} under $10^6$ times channel probings. 
%After the channel probing step, 
%A 
%The BDR  In the simulation, we employ the .  
%As shown in Fig. \ref{fig:BDR}, the BDR decreases with the increase of power $P$, since the negative impacts of noises in uplink and downlink channel probing are both reduced. 
%We also note that deploying an RIS significantly reduce the BDR in PKG system. This is because the RIS establish favorable channel when the remote user, Bob, suffers from poor channel conditions.
%Also, we observe that 
As shown in Fig. \ref{fig:BDR}, the algorithm that designed under the spatially correlated channel model achieves lower BDR than that under the benchmarks. Indeed, the optimized beamforming design improves the ratio of the power of the combined reciprocal component to the noise. 
%These results clearly show the effectiveness of the proposed method in improving the channel reciprocity in RIS-assisted PKG systems. 
%\subsection{Randomness Evaluation}
\begin{table}[htbp]
	\centering  
	\caption{NIST random test result}  
%	\vspace{1em}
	\label{table1}  
	\begin{tabular}{|c|c|c|c|c|c|c|c|c|c|}  
		\hline
		& Pass ratio & P-value\\
		\hline
		Approximate entropy & 0.9833 & 0.4862 \\ \hline
		Runs &  0.9893 & 0.4979 \\ \hline
		Ranking &  0.9901 & 0.4959 \\ \hline
		Longest runs of ones & 0.9899 & 0.4959 \\ \hline
		Frequency & 0.9900 & 0.5047 \\ \hline
		FFT & 0.9853 & 0.4825 \\ \hline
		Block frequency & 0.9887  & 0.5013 \\ \hline
		Cumulative sums & 0.9908 & 0.5197\\ \hline
		Serial & 0.9901 & 0.5275, 0.4917\\\hline
	\end{tabular}
\end{table} 
Finally, we conduct the National Institute of Standards and Technology (NIST) randomness test~\cite{rukhin2000} on the quantized bits to verify the randomness of the obtained bit sequences for cryptographic applications. 
The output of the NIST is called p-value and the tested bits pass the NIST test if the p-value is greater than the threshold 0.01. 
In the simulation, we perform 9 kinds of NIST tests for $10,000$ trials. The test results are shown in Table I, where the pass ratio denotes the ratio of the number of passed trials to the number of all trials. 
From this table, the pass ratio of 9 NIST tests is higher than 0.9 and the average p-value is greater than 0.01. These results indicate the excellent randomness of the bits generated by the proposed method. 
%We performed 9 kinds of NIST tests for 10,000 trials~\cite{22Sum-RIS}, 
%and the ratio of the number of passed trivals to the number of all trivals are high than 0.9. 
%and the average p-value is greater than 0.01. 
%The results indicate good randomness of the bits generated by the proposed method. 

\section{Conclusion}
In this paper, we introduced a transmit and reflective beamforming-based RIS-assisted PKG framework in multi-antenna spatially correlated channels. Based on this framework, we derived the general closed-form KGR expression and formulated an optimization problem to maximize the minimum KGR. We designed a BSUM with a Mirror-Prox algorithm to tackle the non-convex optimization problem. 
Our analysis proved that the KGR increases with the spatial correlation
between the BS antennas and RIS elements. 
In particular, the KGR can be improved significantly with the increase of RIS elements, while it increases with diminishing returns when the number of BS antennas is sufficiently large.
Numerical results showed that our method achieves higher KGR and lower BDR  compared to the benchmarks in the same eavesdropping condition.
%confirmed the  performance of the proposed algorithm and the analysis of the spatial correlation. 

 \appendix

%% appendices
\begin{appendices}
\subsection{Proof of Lemma~1}\label{sec:covariance_calculation}
First, we calculate the covariance of channel estimation $\hat{ h}_{a}$ as 
\begin{align}
	\mathcal{R}_{aa} 
	%	 &=\mathbb{E}\left\{ \tilde{ h}_{a} \tilde{ h}_{a}^H \right\} \\ 
	%	&= P_{B}\mathbb{E}\{ \boldsymbol{w}^T (\boldsymbol{G}_{ra} \boldsymbol{\Phi} \boldsymbol{h}_{br} + \boldsymbol{h}_{ba}  )(\boldsymbol{G}_{ra} \boldsymbol{\Phi} \boldsymbol{h}_{br} + \boldsymbol{h}_{ba}  )^H   \boldsymbol{w}^* \} + P_{A}\sigma_{a}^{ 2 }   \\
	%	&= \mathbb{E}\{ \boldsymbol{w}^T \boldsymbol{G} \boldsymbol{\Phi} \boldsymbol{h}_{br} \boldsymbol{h}_{br}^H \boldsymbol{\Phi}^H \boldsymbol{G}^H \boldsymbol{w}^* +    \boldsymbol{w}^T \boldsymbol{h}_{ba}  \boldsymbol{h}_{ba}^H \boldsymbol{w}^*   \} + P_{A}\sigma_{a}^{ 2 }  \\
	 &=P_\mathrm{B} \boldsymbol{w}^{\sf T} \mathbb{E}\{  \boldsymbol{G}_{ar} \boldsymbol{\Phi} \boldsymbol{h}_{rb} \boldsymbol{h}_{rb}^{\sf H} \boldsymbol{\Phi}^{\sf H} \boldsymbol{G}_{ar}^{\sf H}  \} \boldsymbol{w}^* \notag \\
	 &\quad    +    P_\mathrm{B} \boldsymbol{w}^{\sf T} \mathbb{E}\{\boldsymbol{h}_{ab}  \boldsymbol{h}_{ab}^{\sf H} \} \boldsymbol{w}^* +  ||\boldsymbol{w}||_{2}^{2} \sigma_{a}^{ 2 }. \label{Ra}
\end{align}
By exploting the channel correlations 
%of $ \boldsymbol{G}_{ar} $, $ \boldsymbol{h}_{br} $, and $ \boldsymbol{h}_{ba}$ 
in (\ref{eq1})--(\ref{eq3}), the first term in (\ref{Ra}) is given by
%\begin{align}
%	& \quad \mathbb{E}\{  \boldsymbol{G}_{ra} \boldsymbol{\Phi} \boldsymbol{h}_{br} \boldsymbol{h}_{br}^H \boldsymbol{\Phi}^H \boldsymbol{G}_{ra}^H  \} \\
%	& = \beta_{ r } \mathbb{E}\{    \boldsymbol{R}_{S}^{\frac{1}{2}}\tilde{ \boldsymbol{H}}_{ra} \boldsymbol{R}_{I}^{\frac{1}{2}}     \boldsymbol{\Phi}  \boldsymbol{R}_{I}^{\frac{1}{2}} \tilde{ \boldsymbol{h}}_{br} \tilde{ \boldsymbol{h}}_{br} ^H \boldsymbol{R}_{I}^{\frac{1}{2}}  \boldsymbol{\Phi}^H  \boldsymbol{R}_{I}^{\frac{1}{2}}  \tilde{ \boldsymbol{H}}_{ra}^H \boldsymbol{R}_{S}^{\frac{1}{2}}       \} \\
%	& = \beta_{ r } \boldsymbol{R}_{S}^{\frac{1}{2}} \mathbb{E}\left\{    \left(\sum_{i=1}^{N} \boldsymbol{h}_{i} \boldsymbol{r}_{ i }^{ T } \tilde{ \boldsymbol{h}}_{br} \right)  \left(\sum_{i=1}^{N} \boldsymbol{h}_{i} \boldsymbol{r}_{ i }^{ T } \tilde{ \boldsymbol{h}}_{br} \right)^H
%	\right\} \boldsymbol{R}_{S}^{\frac{1}{2}} \\
%	& = \beta_{ r } \boldsymbol{R}_{S}^{\frac{1}{2}} \mathbb{E}\left\{ \boldsymbol{I}_{M}   \sum_{i=1}^{N} \boldsymbol{r}_{ i }^{ T } \tilde{ \boldsymbol{h}}_{br} \tilde{ \boldsymbol{h}}_{br} ^H \boldsymbol{r}_{ i }^{ * }
%	\right\} \boldsymbol{R}_{S}^{\frac{1}{2}} \\
%	& = \beta_{ r } \boldsymbol{R}_{S}  \sum_{i=1}^{N} \boldsymbol{r}_{ i }^{ T }  \boldsymbol{r}_{ i }^{ * }
%	= \beta_{ r }  || \boldsymbol{R}_{I}^{\frac{1}{2}}     \boldsymbol{\Phi}  \boldsymbol{R}_{I}^{\frac{1}{2}} ||_{F}^{ 2 } \boldsymbol{R}_{S} 
%\end{align}
\begin{align}
	& \quad \mathbb{E}\{  \boldsymbol{G}_{ar} \boldsymbol{\Phi} \boldsymbol{h}_{rb} \boldsymbol{h}_{rb}^{\sf H} \boldsymbol{\Phi}^{\sf H} \boldsymbol{G}_{ar}^{\sf H}  \} \\
	%	&= \beta_{ r }  \boldsymbol{R}_{s} 
	%	\mathbb{E}\{    \tilde{ \boldsymbol{H}}_{ar} \boldsymbol{R}_{I}^{\frac{1}{2}}     \boldsymbol{\Phi}  \boldsymbol{R}_{I}^{\frac{1}{2}} \tilde{ \boldsymbol{h}}_{br} \tilde{ \boldsymbol{h}}_{br} ^H \boldsymbol{R}_{I}^{\frac{1}{2}}  \boldsymbol{\Phi}^H  \boldsymbol{R}_{I}^{\frac{1}{2}}  \tilde{ \boldsymbol{H}}_{ar}^H      \}  \\
	%	& = \frac{\beta_{ r }}{ M }  \boldsymbol{R}_{s}  
	%	\mathbb{E}\{\tilde{ \boldsymbol{h}}_{br} ^H \boldsymbol{R}_{I}^{\frac{1}{2}}  \boldsymbol{\Phi}^H  \boldsymbol{R}_{I}^{\frac{1}{2}}  \tilde{ \boldsymbol{H}}_{ar}^H \tilde{ \boldsymbol{H}}_{ar} \boldsymbol{R}_{I}^{\frac{1}{2}}     \boldsymbol{\Phi}  \boldsymbol{R}_{I}^{\frac{1}{2}} \tilde{ \boldsymbol{h}}_{br}          \}  \\
	%	& =   \beta_{ r }  \boldsymbol{R}_{s}  
	%	\mathbb{E}\{\tilde{ \boldsymbol{h}}_{br} ^H \boldsymbol{R}_{I}^{\frac{1}{2}}  \boldsymbol{\Phi}^H  \boldsymbol{R}_{I}^{\frac{1}{2}}   \boldsymbol{R}_{I}^{\frac{1}{2}}     \boldsymbol{\Phi}  \boldsymbol{R}_{I}^{\frac{1}{2}} \tilde{ \boldsymbol{h}}_{br}          \}   \\ 
	& = \beta_{ r }  \boldsymbol{R}_{S}  
	\mathbb{E}\{  \text{vec}\{\tilde{ \boldsymbol{h}}_{rb} ^{\sf H} \boldsymbol{R}_{I}^{\frac{1}{2}}  \boldsymbol{\Phi}^{\sf H}  \boldsymbol{R}_{I}^{\frac{1}{2}} \}^{\sf H}    \text{vec}\{\tilde{ \boldsymbol{h}}_{rb} ^{\sf H} \boldsymbol{R}_{I}^{\frac{1}{2}}  \boldsymbol{\Phi}^{\sf H}  \boldsymbol{R}_{I}^{\frac{1}{2}} \}        \}  \\   
	& \overset{(a)}{=} \beta_{ r }  \boldsymbol{R}_{S}  
	\mathbb{E}\{   \boldsymbol{v}^{\sf H}   ( (\boldsymbol{R}_{I}^{\frac{1}{2}} )^{\sf T} \odot  (\boldsymbol{R}_{I}^{\frac{1}{2}} ) )^{\sf H} (\tilde{ \boldsymbol{h}}_{rb} ^* \otimes \boldsymbol{I}_{N}   )          \notag \\
	&\quad \times(\tilde{ \boldsymbol{h}}_{rb} ^{\sf T} \otimes \boldsymbol{I}_{N}   )  ( (\boldsymbol{R}_{I}^{\frac{1}{2}} )^{\sf T} \odot  (\boldsymbol{R}_{I}^{\frac{1}{2}} ) )  \boldsymbol{v}     \}  \\  
	& = \beta_{ r }  \boldsymbol{R}_{S}  
	\boldsymbol{v}^{\sf H}   ( (\boldsymbol{R}_{I}^{\frac{1}{2}} )^{\sf T} \odot  (\boldsymbol{R}_{I}^{\frac{1}{2}} ) )^{\sf H}       ( (\boldsymbol{R}_{I}^{\frac{1}{2}} )^{\sf T} \odot  (\boldsymbol{R}_{I}^{\frac{1}{2}} ) )  \boldsymbol{v}      
	\notag \\
	&= \beta_{ r }  \boldsymbol{R}_{s}  
	\boldsymbol{v}^{\sf H}   ( \boldsymbol{R}_{I} ^{\sf T } \circ  {\boldsymbol{R}}_{I} ) \boldsymbol{v}  
	= \beta_{ r }  \boldsymbol{R}_{S}  
	\boldsymbol{v}^{\sf H}  \tilde{\boldsymbol{R}}_{I}         \boldsymbol{v}.
\end{align}
$(a)$ follows from $ \text{vec}( \A \B \C ) = (\C^{\sf T} \otimes \A) \text{vec}(\B)$ and $ \text{vec}( \A \diag(\d) \C ) = (\C^{\sf T} \odot \A) \d$. 
Then, the second term of the right-hand side of (\ref{Ra}) is calculated as $\mathbb{E}\{\boldsymbol{h}_{ab}  \boldsymbol{h}_{ab}^{\sf H} \} = \beta_{ ab } \boldsymbol{R}_{S}$.
%At the same time, we can obtain that 
%\begin{align}
%	\boldsymbol{w}^T \mathbb{E}\{\boldsymbol{h}_{ba}  \boldsymbol{h}_{ba}^H \} \boldsymbol{w}^{ * }= \beta_{d} \boldsymbol{w}^T \boldsymbol{R}_{S} \boldsymbol{w}^{ * }
%\end{align}
%Thus, the covariance is obtained as
%\begin{align}
%	\mathcal{R}_{aa} &= P_{\mathrm{B}}\beta_{ r } \boldsymbol{w}^T  \boldsymbol{R}_{S}  \boldsymbol{w}^{ * }
%	\boldsymbol{v}^H   ( \boldsymbol{R}_{I} ^{ T } \circ   \boldsymbol{R}_{I})           \boldsymbol{v} \notag \\
%	& \quad + P_{\mathrm{B}}\beta_{d} \boldsymbol{w}^T \boldsymbol{R}_{S} \boldsymbol{w}^{ * } + ||\boldsymbol{w}||^2 \sigma_{ a } ^ { 2 }
%\end{align}
Similarly, we have
%the other covariances can be calculated as 
\begin{align}
	\mathcal{R}_{bb} &= \beta_{ r } \boldsymbol{w}^{\sf T}  \boldsymbol{R}_{S}  \boldsymbol{w}^{ * }
	\boldsymbol{v}^{\sf H}   \tilde{\boldsymbol{R}}_{I}           \boldsymbol{v}  + \beta_{ab} \boldsymbol{w}^{\sf T} \boldsymbol{R}_{S} \boldsymbol{w}^{ * } +  \sigma_{ b } ^ { 2 }, \\
	\mathcal{R}_{ab}&=\sqrt{P_{\mathrm{B}}}\beta_{ r } \boldsymbol{w}^{\sf T}  \boldsymbol{R}_{S}  \boldsymbol{w}^{ * }
	\boldsymbol{v}^{\sf H}   \tilde{\boldsymbol{R}}_{I}          \boldsymbol{v}  + \sqrt{P_{\mathrm{B}}}\beta_{ab} \boldsymbol{w}^{\sf T} \boldsymbol{R}_{S} \boldsymbol{w}^{ * } \notag \\ &=\mathcal{R}_{ba},  \\
	% ek
	\mathcal{R}_{e_{k}e_{k}} &= \beta_{ r }^{k} \boldsymbol{w}^{\sf T}  \boldsymbol{R}_{S}  \boldsymbol{w}^{ * }
	\boldsymbol{v}^{\sf H}   \tilde{\boldsymbol{R}}_{I}          \boldsymbol{v}  + \beta_{ak} \boldsymbol{w}^{\sf T} \boldsymbol{R}_{S} \boldsymbol{w}^{ * } +  \sigma_{ e_{k} } ^ { 2 }, \\%R_{ejej}
	\mathcal{R}_{be_{k}} &=\sqrt{\beta_{ r }\beta_{ r }^{k}} 
	\boldsymbol{v}^{\sf H}   \left[ ((\boldsymbol{R}_{I}^{\frac{1}{2}}) ^{\sf T }\boldsymbol{R}_{k}^{r}(\boldsymbol{R}_{I}^{\frac{1}{2}}) ^{\sf T }) \circ   \boldsymbol{R}_{I}\right]           \boldsymbol{v}    \boldsymbol{w}^{\sf T}  \boldsymbol{R}_{S}  \boldsymbol{w}^{ * } \notag \\ 
	&\quad +  \sqrt{\beta_{ ab }\beta_{ ak }} \boldsymbol{w}^{\sf T} \boldsymbol{R}_{S}^{\frac{1}{2}} \boldsymbol{R}_{k}^{d}\boldsymbol{R}_{S}^{\frac{1}{2}} \boldsymbol{w}^{ * } \notag \\ &=\mathcal{R}_{e_{k}b}^{\sf H} = \mathcal{R}_{ae_{k}}/\sqrt{P_{\rm B}}= \mathcal{R}_{e_{k}a}^{\sf H}/\sqrt{P_{\rm B}}.
%	,   \\%%
%	\mathcal{R}_{ae_{k}} &= \mathcal{R}_{e_{k}a}^{\sf H} = \sqrt{P_{\mathrm{B}}}	\mathcal{R}_{be_{k}}.
\end{align}
Then, we can calculate the two determinants in (\ref{eq:MI}) and obtain the KGR as (\ref{eq:whole}).

\subsection{Proof of Lemma~\ref{lemma-conver}}\label{appendix-conver}
%\begin{IEEEproof}
	We denote 
	the objective value of Problem (\ref{eq:real}) as $\varphi(\tilde{\boldsymbol{w}},\tilde{\boldsymbol{v}}) = \min_{k} \tilde{f}_{k}(\tilde{\boldsymbol{w}}, \tilde{\boldsymbol{v}})$.
	By denoting the objective value of the subproblem in (\ref{eq:subproblem-v}) and the subproblem in (\ref{eq:subproblem-w})  as $\bar{\varphi}(\tilde{\boldsymbol{w}},\tilde{\boldsymbol{v}})$ and $\tilde{\varphi}(\tilde{\boldsymbol{w}},\tilde{\boldsymbol{v}})$, respectively, we have 
%	\begin{align}
	$	\varphi(\tilde{\boldsymbol{w}}^{(i+1)},\tilde{\boldsymbol{v}}^{(i+1)})
	 	 \overset{(b)}{\geq} \bar{\varphi}(\tilde{\boldsymbol{w}}^{(i+1)},\tilde{\boldsymbol{v}}^{(i+1)})\overset{(c)}{=} \tilde{\varphi}(\tilde{\boldsymbol{w}}^{(i+1)},\tilde{\boldsymbol{v}}^{(i+1)})  \overset{(d)}{\geq} \tilde{\varphi}(\tilde{\boldsymbol{w}}^{(i)},\tilde{\boldsymbol{v}}^{(i+1)})$,
%	 \end{align}
	where inequality ($b$) holds because $\bar{\varphi}(\tilde{\boldsymbol{w}},\tilde{\boldsymbol{v}})$ is a lower bound of $\varphi(\tilde{\boldsymbol{w}},\tilde{\boldsymbol{v}})$,  inequality ($c$) holds since $\tilde{\varphi}(\tilde{\boldsymbol{w}},\tilde{\boldsymbol{v}})$ is a local approximation of $\bar{\varphi}(\tilde{\boldsymbol{w}},\tilde{\boldsymbol{v}})$ at the point $(\tilde{\boldsymbol{w}}^{(i+1)},\tilde{\boldsymbol{v}}^{(i+1)})$. The inequality ($d$) holds since $ \tilde{\boldsymbol{w}}^{(i+1)}= \arg \max_{\tilde{\boldsymbol{w}}}  \tilde{\varphi}(\tilde{\boldsymbol{w}},\tilde{\boldsymbol{v}}^{(i+1)})$.
	Similarly, we have 
%	\begin{align}
$
		\tilde{\varphi}(\tilde{\boldsymbol{w}}^{(i)},\tilde{\boldsymbol{v}}^{(i+1)}) \overset{(e)}{\geq} \bar{\varphi}(\tilde{\boldsymbol{w}}^{(i)},\tilde{\boldsymbol{v}}^{(i+1)}) \overset{(f)}{\geq} \bar{\varphi}(\tilde{\boldsymbol{w}}^{(i)},\tilde{\boldsymbol{v}}^{(i)})
		 \overset{(g)}{=} \varphi(\tilde{\boldsymbol{w}}^{(i)},\tilde{\boldsymbol{v}}^{(i)})$,
%	\end{align}
where ($e$) holds since $\bar{\varphi}(\tilde{\boldsymbol{w}},\tilde{\boldsymbol{v}}^{(i+1)})$ is an lower bound of $\tilde{\varphi}(\tilde{\boldsymbol{w}},\tilde{\boldsymbol{v}}^{(i+1)})$,  ($f$) holds because $\tilde{\boldsymbol{v}}^{(i+1)}  = \arg \max_{\tilde{\boldsymbol{v}}} \bar{\varphi}(\tilde{\boldsymbol{w}}^{(i)},\tilde{\boldsymbol{v}})$, and ($g$) holds since $\bar{\varphi}(\tilde{\boldsymbol{w}},\tilde{\boldsymbol{v}})$ is a local approximation of $\varphi(\tilde{\boldsymbol{w}},\tilde{\boldsymbol{v}})$ at the point $(\tilde{\boldsymbol{w}}^{(i)},\tilde{\boldsymbol{v}}^{(i)})$.
%	Also, since $\bar{R}_{(t)}(\tilde{\boldsymbol{w}},\tilde{\boldsymbol{v}})$ is the lower bound of $R_{(t)}(\tilde{\boldsymbol{w}},\tilde{\boldsymbol{v}}_{(t)}^{(i+1)})$ and $\bar{R}_{(t)}(\tilde{\boldsymbol{w}},\tilde{\boldsymbol{v}})$ is the lower bound of $R_{(t)}(\tilde{\boldsymbol{w}},\tilde{\boldsymbol{v}})$, we have
%	\begin{align}
%		R_{(t)}(\tilde{\boldsymbol{w}}_{(t)}^{(i+1)},\tilde{\boldsymbol{v}}_{(t)}^{(i+1)}) \geq \tilde{R}_{(t)}(\tilde{\boldsymbol{w}}_{(t)},\tilde{\boldsymbol{v}}_{(t)}^{(i+1)})
%	\end{align}
%	Also, 
Therefore, we have proved the equality $\varphi(\tilde{\boldsymbol{w}}^{(i+1)},\tilde{\boldsymbol{v}}^{(i+1)}) \geq \varphi(\tilde{\boldsymbol{w}}^{(i)},\tilde{\boldsymbol{v}}^{(i)}) $, i.e., the sequence $\{\varphi(\tilde{\boldsymbol{w}}^{(i)},\tilde{\boldsymbol{v}}^{(i)})\}_{i=0}^{\infty}$ is non-increasing over iterations. 
%	According to the BMM algorithm \cite{razaviyayn2013unified}, we have
%	\begin{align}
%		\min_{k} f_{k}^{(i)}(\tilde{\boldsymbol{w}}^{(t+1)},\tilde{\boldsymbol{v}}^{(t+1)}) &= \min_{k} \hat{f}_{k}^{(i)}(\tilde{\boldsymbol{w}}^{(t+1)},\tilde{\boldsymbol{v}}^{(t+1)})\\
%		&\geq \min_{k} l_{k}^{(i)}(\tilde{\boldsymbol{w}}^{(t)},\tilde{\boldsymbol{v}}^{(t+1)})\\
%		&\geq \min_{k} l_{k}^{(i)}(\tilde{\boldsymbol{w}}^{(t)},\tilde{\boldsymbol{v}}^{(t)})\\
%		&= \min_{k} f_{k}^{(i)}(\tilde{\boldsymbol{w}}^{(t)},\tilde{\boldsymbol{v}}^{(t)})
%	\end{align}
	In addition, assuming the channel gain is finite and the transmit power $P_{\rm A}$ and $P_{\rm B}$ is limited, the solution set is compact and the 
	 KGR has an upper bound. Consequently, the objective values of Problem (\ref{eq:real})
%	the sequence of iterations  $\{\tilde{\boldsymbol{v}}^{(i)},\tilde{\boldsymbol{w}}^{(i)}\}_{i=1}^{\infty}$ 
	are non-increasing and convergence. 	

	\subsection{Proof of Lemma~\ref{lemma-L}}\label{appendix-L}
%\begin{IEEEproof}
	According to \cite{Mirror-Prox}, the function $\Psi^{(i)}(\cdot)$ is monotone and Lipschitz continuous. Then, $\Psi^{(i)}(\cdot)$ is defined as $L$-Lipschitz
	if it satisfies with the following constraints \cite{fastfpp}
	\begin{align}
		&\left\|\nabla_{\tilde{\boldsymbol{v}}} \psi^{(i)}(\tilde{\boldsymbol{v}}, \boldsymbol{y})-\nabla_{\tilde{\boldsymbol{v}}} \psi^{(i)}\left(\tilde{\boldsymbol{v}}^{\prime}, \boldsymbol{y}\right)\right\|_{2} \leq L \left\|\tilde{\boldsymbol{v}}-\tilde{\boldsymbol{v}}^{\prime}\right\|_{2}, \label{eq:a}\\
		&\left\|\nabla_{\boldsymbol{y}} \psi^{(i)}(\tilde{\boldsymbol{v}}, \boldsymbol{y})-\nabla_{\boldsymbol{y}} \psi^{(i)}\left(\tilde{\boldsymbol{v}}, \boldsymbol{y}^{\prime}\right)\right\|_{\infty} \leq L \left\|\boldsymbol{y}-\boldsymbol{y}^{\prime}\right\|_{1}, \label{eq:b}\\
		&\left\|\nabla_{\tilde{\boldsymbol{v}}} \psi^{(i)}(\tilde{\boldsymbol{v}}, \boldsymbol{y})-\nabla_{\tilde{\boldsymbol{v}}} \psi_{(t)}^{(i)}\left(\tilde{\boldsymbol{v}}, \boldsymbol{y}^{\prime}\right)\right\|_{2} \leq L \left\|\boldsymbol{y}-\boldsymbol{y}^{\prime}\right\|_{1}, \label{eq:c}\\
		&\left\|\nabla_{\boldsymbol{y}} \psi^{(i)}(\tilde{\boldsymbol{v}}, \boldsymbol{y})-\nabla_{\boldsymbol{y}} \psi^{(i)}\left(\tilde{\boldsymbol{v}}^{\prime}, \boldsymbol{y}\right)\right\|_{\infty} \leq L \left\|\tilde{\boldsymbol{v}}-\tilde{\boldsymbol{v}}^{\prime}\right\|_{2}. \label{eq:d}
	\end{align}
%	where $\|\cdot\|_{\mathcal{V}}$ and $\|\cdot\|_{\mathcal{Y}}$ are the norms embeded in $\mathcal{V}$ and $\mathcal{Y}$, respectively,
%	$\|\cdot\|_{\mathcal{V}}^{*}$ and $\|\cdot\|_{\mathcal{Y}}^{*}$ are the dual norms of $\|\cdot\|_{\mathcal{V}}$ and $\|\cdot\|_{\mathcal{Y}}$, respectively. In problem $(\ref{eq:mirror-v})$, the  $\|\cdot\|_{\mathcal{V}}$ and $\|\cdot\|_{\mathcal{Y}}$ are the $\ell_{2}$ norm and $\ell_{1}$ norm, respectively. Hence the $\|\cdot\|_{\mathcal{V}}^{*}$ and $\|\cdot\|_{\mathcal{Y}}^{*}$ are the $\ell_{2}$ norm and $\ell_{\infty}$ norms, respectively. 
	First, the equality in (\ref{eq:a}) holds because
%	\begin{align}
		$\left\|\nabla_{\tilde{\boldsymbol{v}}} \psi^{(i)}(\tilde{\boldsymbol{v}}, \boldsymbol{y})-\nabla_{\tilde{\boldsymbol{v}}} \psi^{(i)}\left(\tilde{\boldsymbol{v}}^{\prime}, \boldsymbol{y}\right)\right\|_{2} 
%		&= \|2 (\bar{\boldsymbol{\tau}}^{(i)})^{\sf T}\boldsymbol{y} \tilde{\boldsymbol{v}} - 
%		2 (\bar{\boldsymbol{\tau}}^{(i)})^{\sf T}\boldsymbol{y} \tilde{\boldsymbol{v}}^{\prime}
%		\|_{2}   \label{ineq:1-2}\\
		\leq \|2 (\bar{\boldsymbol{\tau}}^{(i)})^{\sf T}\boldsymbol{y}\|_{2} \|\tilde{\boldsymbol{v}} - \tilde{\boldsymbol{v}}^{\prime}\|_{2} 
%		 \label{ineq:1-3}\\
%		& \leq  2\|\bar{\boldsymbol{\tau}}^{(i)}\|_{2} \|\boldsymbol{y}\|_{2} \|\tilde{\boldsymbol{v}} - \tilde{\boldsymbol{v}}^{\prime}\|_{2} \label{ineq:1-4} \\
		 \leq L\|\tilde{\boldsymbol{v}} - \tilde{\boldsymbol{v}}^{\prime}\|_{2}$,
%		 \label{ineq:1-5}
%	\end{align}
%	where 
%	(\ref{ineq:1-3})  follows the Cauchy-Schwarz inequality \cite{Cauchy}, 
%	 (\ref{ineq:1-5}) 
	 which follows
	 the Cauchy-Schwarz inequality. 
%	 \cite{Cauchy}. 
%	in problem (\ref{eq:subproblem-v}), the $\boldsymbol{y}$ satisfies $0 \leq y_{k} \leq 1, k\in\{1,\cdots,K\}$, and thus
%	 $\|\boldsymbol{y}\|_{2} \leq \|\boldsymbol{y}\|_{1} = 1$.
	Then, the equality in (\ref{eq:b}) holds since
%	\begin{align}
		$\nabla_{\boldsymbol{y}}\psi^{(i)}(\tilde{\boldsymbol{v}},\boldsymbol{y})=	\nabla_{\boldsymbol{y}}\psi^{(i)}(\tilde{\boldsymbol{v}},\boldsymbol{y}^{\prime}) =  \bar{\boldsymbol{\tau}}^{(i)} \tilde{\boldsymbol{v}}^{\top} \tilde{\boldsymbol{v}}  + 
		\mathbf{P}^{(i)} \tilde{\boldsymbol{v}}+\mathbf{q}^{(i)}$,
%	\end{align}
	which is irrelevant to the variable $\boldsymbol{y}$.
	In addition, the inequality in (\ref{eq:c}) holds due to
	\begin{align}
		&\quad \left\|\nabla_{\tilde{\boldsymbol{v}}} \psi^{(i)}(\tilde{\boldsymbol{v}}, \boldsymbol{y})-\nabla_{\tilde{\boldsymbol{v}}} \psi^{(i)}\left(\tilde{\boldsymbol{v}}, \boldsymbol{y}^{\prime}\right)\right\|_{2} \\
%		& = \|2 (\bar{\boldsymbol{\tau}}^{(i)})^{\sf T}\boldsymbol{y}\tilde{\boldsymbol{v}} + \P^{\sf T} \boldsymbol{y} - 2(\bar{\boldsymbol{\tau}}^{(i)})^{\sf T}\boldsymbol{y}^{\prime}\tilde{\boldsymbol{v}} - \P^{\sf T} \boldsymbol{y}^{\prime} \|_{2}  \\
		& \leq 2 \|(\bar{\boldsymbol{\tau}}^{(i)})^{\sf T}\boldsymbol{y} - (\bar{\boldsymbol{\tau}}^{(i)})^{\sf T}\boldsymbol{y}^{\prime}\|_{2} \|\tilde{\boldsymbol{v}}\|_{2} + \|\P^{\sf T}\boldsymbol{y} - \P^{\sf T}\boldsymbol{y}^{\prime}\|_{2} \label{ineq:triangle}\\
		& \leq 2 \left(\|\bar{\boldsymbol{\tau}}^{(i)}\|_{2}   \sqrt{N} + \left( \max_{k}\  \|\p_{k}^{(i)}\|_{2} \right)\right)\|\boldsymbol{y} - \boldsymbol{y}^{\prime}\|_{1} \\
%		&= \left(2\sqrt{N} \|\bar{\boldsymbol{\tau}}^{(i)}\|_{2} +  \max_{k}\  \|\p_{k}^{(i)}\|_{2}  \right)\|\boldsymbol{y} - \boldsymbol{y}^{\prime}\|_{1}  \\
		 &= L\|\boldsymbol{y} - \boldsymbol{y}^{\prime}\|_{1},
	\end{align}
where (\ref{ineq:triangle}) follows the triangle inequality.  
%\cite{maligranda2008some}. 
At last, the inequality (\ref{eq:d}) holds as
	\begin{align}
		&\quad \left\|\nabla_{\boldsymbol{y}} \psi^{(i)}(\tilde{\boldsymbol{v}}, \boldsymbol{y})-\nabla_{\boldsymbol{y}} \psi^{(i)}\left(\tilde{\boldsymbol{v}}^{\prime}, \boldsymbol{y}\right)\right\|_{\infty}  \notag \\
%		&=\| \bar{\boldsymbol{\tau}}^{(i)} \tilde{\boldsymbol{v}}^{\top} \tilde{\boldsymbol{v}}  + 
%		\mathbf{P}^{(i)} \tilde{\boldsymbol{v}} - 
%		\bar{\boldsymbol{\tau}}^{(i)} (\tilde{\boldsymbol{v}}^{\prime})^{\top} \tilde{\boldsymbol{v}}^{\prime}  -
%		\mathbf{P}^{(i)} \tilde{\boldsymbol{v}}^{\prime} \|_{\infty} \\
		& \leq \| \bar{\boldsymbol{\tau}}^{(i)} \tilde{\boldsymbol{v}}^{\top} \tilde{\boldsymbol{v}}
		- 
		\bar{\boldsymbol{\tau}}^{(i)} (\tilde{\boldsymbol{v}}^{\prime})^{\top} \tilde{\boldsymbol{v}}^{\prime} \|_{\infty} +  
		\| 
		\mathbf{P}^{(i)} \tilde{\boldsymbol{v}} - 
		\mathbf{P}^{(i)} \tilde{\boldsymbol{v}}^{\prime} \|_{\infty} \\
		& \leq \|\bar{\boldsymbol{\tau}}^{(i)}\|_{\infty} \| \tilde{\boldsymbol{v}} + \tilde{\boldsymbol{v}}^{\prime} \|_{\infty}\| \tilde{\boldsymbol{v}} - \tilde{\boldsymbol{v}}^{\prime} \|_{\infty} + \max_{k}\left\{ |(\p_{k}^{(i)})^{\sf T}(\tilde{\boldsymbol{v}} - \tilde{\boldsymbol{v}}^\prime)  |  \right\} \\
		& \leq \left(\|\bar{\boldsymbol{\tau}}^{(i)}\|_{2}
		(\|\tilde{\boldsymbol{v}}\|_{\infty} + \|\tilde{\boldsymbol{v}}^{\prime}\|_{\infty}) + \max_{k}\left\{ \|(\p_{k}^{(i)}) \|_{2}  \right\}\right) \|(\tilde{\boldsymbol{v}} - \tilde{\boldsymbol{v}}^\prime)\|_{2}\\  
		 &= L \|(\tilde{\boldsymbol{v}} - \tilde{\boldsymbol{v}}^\prime)\|_{2}.
	\end{align}

\end{appendices}

\bibliographystyle{IEEEtran}
\bibliography{IEEEabrv,Ref1}

\end{document}